\documentclass[times,twocolumn,final]{elsarticle}
\usepackage{framed,multirow}

\usepackage{amssymb}
\usepackage{latexsym}

\usepackage{url}
\usepackage{xcolor}

\usepackage{hyperref}
\usepackage{bm}
\usepackage{makecell}
\usepackage{graphicx}
\definecolor{newcolor}{rgb}{.8,.349,.1}

\begin{document}

%\verso{Given-name Surname \textit{et~al.}}

\begin{frontmatter}

% \title{Beyond Strong labels: Weakly-supervised Learning Based on Gaussian Pseudo Labels for The Segmentation of Ellipse-like Vascular Structures in Non-contrast CTs\tnoteref{tnote1}}%
% \tnotetext[tnote1]{This is an example for title footnote coding.}
\title{Beyond Strong labels: Weakly-supervised Learning Based on Gaussian Pseudo Labels for The Segmentation of Ellipse-like Vascular Structures in Non-contrast CTs}%

% \author[1,2]{Qixiang \snm{Ma}\corref{cor1}}
% \cortext[cor1]{Corresponding author: qixiang.ma@etudiant.univ-rennes1.fr (Qixiang MA)}
% \author[1,2]{Adrien \snm{Kaladji}}
% \author[2,3]{Huazhong \snm{Shu}}
% \author[2,3]{Guanyu \snm{Yang}}
% % \fntext[fn1]{This is author footnote for second author.}
% \author[1,2]{Antoine \snm{Lucas}}
% \author[1,2]{Pascal \snm{Haigron}}

% \address[1]{Univ Rennes, CHU Rennes, Inserm, LTSI – UMR 1099, F-35000 Rennes, France}
% \address[2]{Centre de Recherche en Information Biomédicale Sino-français (CRIBs), Univ Rennes, Inserm, Southeast University, F-35000 Rennes, France, Nanjing 210096, China}
% \address[3]{Laboratory of Image Science and Technology, Southeast University, Nanjing 210096, China}

\author[a1,a2]{Qixiang Ma\corref{cor1}}
\author[a1,a2]{Adrien Kaladji}
\author[a1,a3]{Huazhong Shu}
\author[a1,a3]{Guanyu Yang}
\author[a1,a2]{Antoine Lucas}
\author[a1,a2]{Pascal Haigron}
\address[a1]{Univ Rennes, CHU Rennes, Inserm, LTSI – UMR 1099, F-35000 Rennes, France}
\address[a2]{Centre de Recherche en Information Biomédicale Sino-français (CRIBs), Univ Rennes, Inserm, Southeast University, F-35000 Rennes, France, Nanjing 210096, China}
\address[a3]{Laboratory of Image Science and Technology, Southeast University, Nanjing 210096, China}

\cortext[cor1]{Corresponding author: qixiang.ma@etudiant.univ-rennes1.fr (Qixiang MA)}
% \makeatletter 
% \let\emailauthor
% \@gobbletwo 
% \makeatother
% \emailauthor{JLY@seu.edu.cn}{L. Jiang}
%\received{1 May 2013}
%\finalform{10 May 2013}
%\accepted{13 May 2013}
%\availableonline{15 May 2013}
%\communicated{S. Sarkar}

\begin{abstract}
Deep learning-based automated segmentation of vascular structures in preoperative CT angiography (CTA) images contributes to computer-assisted diagnosis and interventions. While CTA is the common standard, non-contrast CT imaging has the advantage of avoiding complications associated with contrast agents. However, the challenges of labor-intensive labeling and high labeling variability due to the ambiguity of vascular boundaries hinder conventional strong-label-based, fully-supervised learning in non-contrast CTs. This paper introduces a novel weakly-supervised framework using the elliptical topology nature of vascular structures in CT slices. It includes an efficient annotation process based on our proposed standards, an approach of generating 2D Gaussian heatmaps serving as pseudo labels, and a training process through a combination of voxel reconstruction loss and distribution loss with the pseudo labels. We assess the effectiveness of the proposed method on one local and two public datasets comprising non-contrast CT scans, particularly focusing on the abdominal aorta. On the local dataset, our weakly-supervised learning approach based on pseudo labels outperforms strong-label-based fully-supervised learning (1.54\% of Dice score on average), reducing labeling time by around 82.0\%. The efficiency in generating pseudo labels allows the inclusion of label-agnostic external data in the training set, leading to an additional improvement in performance (2.74\% of Dice score on average) with a reduction of 66.3\% labeling time, where the labeling time remains considerably less than that of strong labels. On the public dataset, the pseudo labels achieve an overall improvement of 1.95\% in Dice score for 2D models with a reduction of 68\% of the Hausdorff distance for 3D model.
\end{abstract}


\end{frontmatter}

%\linenumbers

\section{Introduction}
Computed tomography (CT) is a crucial medical imaging modality for visualization and analysis of anatomical structures. The progress of CT over the past decades has led to diverse clinical uses \cite{power2016computed}. One prominent application is CT angiography (CTA), which entails contrast agent injection to enhance luminal density, accentuating the contrast between vascular structures (VS) and surrounding tissues. CTA is routinely employed to enhance visualization of VS \cite{foley2003computed, sun2012coronary}. It aids in diagnosing and assessing vascular conditions such as atherosclerosis, stenosis, aneurysms, and abnormal vessel formations.

Although CTA may serve as the unique approach for discernment of VS to facilitate diagnosis and intervention planning of vascular diseases, it involves several considerable adverse effects \cite{mcdonald2013intravenous, davenport2013contrast, hinson2017risk}. One of the primary concerns associated with CTA is its potential to induce renal complications, particularly in patients with compromised renal function. The intravenous administration of contrast agents, often required for optimal vascular imaging, can strain the kidneys and may lead to contrast-induced nephropathy (CIN) or acute kidney injury (AKI). In addition to renal complications, other considerations include potential allergic reactions to iodine contrast agents and potential harm from needle punctures. Allergic reactions to contrast agents can range from mild to severe, requiring immediate medical attention. Needle punctures for contrast injection could lead to localized discomfort, bruising, or infection. To mitigate these adverse effects, careful patient assessment of renal function and allergies prior to CTA should be performed, which may prompt further investigations, causing patients' apprehension and financial implications. As such, an alternative contrast agent-free CT imaging modality is supposed to be seriously considered.

Non-contrast CT imaging, which omits the use of contrast agents, offers a means to circumvent the risks associated with renal complications, allergies, and injection-related issues during the diagnosis and intervention planning of vascular diseases. Recent research highlights its applicability in context of abdominal aorta diseases. For instance, Kaladji et al. \cite{kaladji2015safety} stated the safe and accurate performance of endovascular aneurysm repair (EVAR) guided by non-contrast CTs in patients who suffer from abdominal aortic aneurysm (AAA). In their study, abdominal aortas with aneurysms in 3D non-contrast CT volumes were manually segmented and then virtually overlaid onto 2D fluoroscopic images to guide minimally invasive procedures. Ma et al. \cite{ma2023deep} further automated the segmentation process using Deep Learning (DL) techniques to facilitate the virtual enhancement of non-contrast cardiovascular CT images.

\begin{figure}[!t]
\centering
% \begin{center}{
\begin{minipage}[c]{1.0\linewidth}
\includegraphics[width=\textwidth]{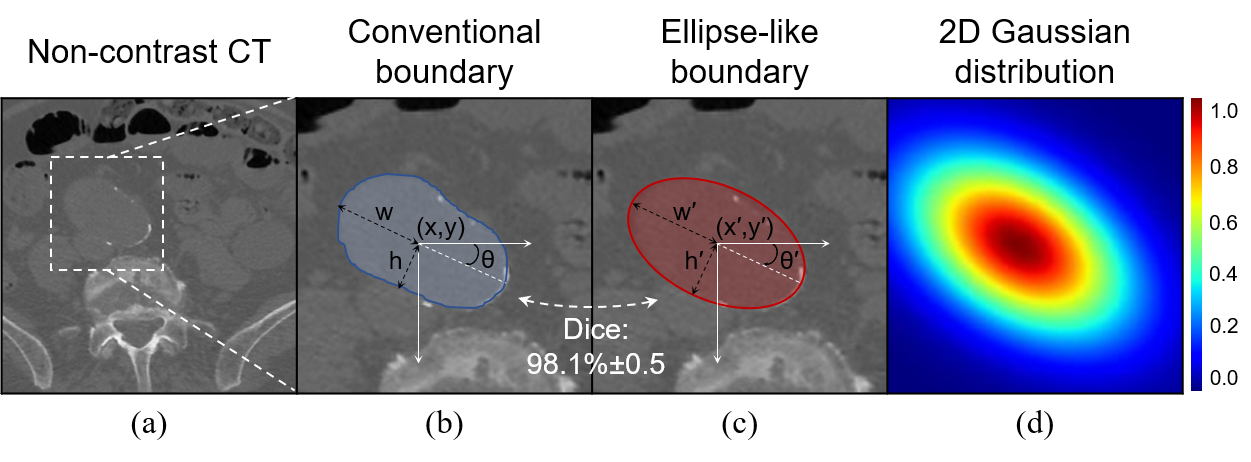}
\end{minipage}
% }\end{center}
\caption{Conventional and elliptical boundaries ((b) and (c)) outlining an aorta derived from a non-contrast CT scan (a). Both enclosed regions share fundamental attributes such as the central point $(x,y)$ and $(x',y')$, rotation angles $\theta$ and $\theta'$, semi-major axes $w$ and $w'$, and semi-minor axes $h$ and $h'$. The computed Dice coefficient for the two enclosed regions is 98.1\%$\pm$0.5 across the entire local dataset. A 2D Gaussian distribution generated from (c) is depicted in (d), containing pixel intensities within the range of $[0.0,1.0]$. This Gaussian distribution functions as a pseudo label.}\label{fig:s1_1}
\end{figure}

The DL models utilize multi-layer architectures to learn representations from complex features, supplanting handcrafted patterns \cite{lecun2015deep}, thus establishing new benchmarks in medical image segmentation \cite{litjens2017survey, minaee2021image}. Despite the effectiveness of deep learning in segmenting VS in non-contrast CT images, two significant challenges persist. Firstly, the data annotation is labor-intensive and time-consuming. The requirement of strong labels (ground truths) often costs a major expenditure of time and effort of multiple experts and the supervision of the surgeon. Secondly, due to the inherent characteristics of non-contrast CTs, as exampled in Figure~\ref{fig:s1_1} (a), the boundaries of the VS are often ambiguous in slices, affecting the accuracy of labeling, increasing intra- and inter-observer variability \cite{ma2023deep}, thereby impacting the precision and stability of training. Emerging weakly-supervised learning approaches recently offered a novel perspective that mitigated annotation costs by utilizing pseudo labels \cite{tajbakhsh2020embracing}. However, this often comes at the expense of sacrificing model accuracy. It is reasonable to assert that accuracy is a pivotal metric in clinical tasks, given its direct impact on diagnosis and treatment decisions. High accuracy ensures reliable segmentation results, thus providing more precise information for clinical decisions. However, the current weakly-supervised learning methods often achieve improved labeling efficiency at the cost of sacrificing accuracy. They struggle to simultaneously achieve advancements in both reducing annotation time and enhancing accuracy.

To address the dual challenge of reducing labor-intensive data labeling while maintaining or even improving the segmentation performance, we present a novel approach for weakly-supervised learning of VS in non-contrast CT. Our observations underscore the ellipse-like topology commonly exhibited by VS in CT slices. Therefore, we argue that pseudo labels, containing representations of these elliptical structures, can serve as substitutes for traditional strong labels. These pseudo labels capture essential features of the VS, including topology, position, and orientation, thus potentially affording enhanced training benefits to deep learning models compared to their strong label counterparts due to the explicit nature of the topology. As an instance, in Figure~\ref{fig:s1_1}, the area enclosed by an elliptical boundary (c) maintains fundamental characteristics and exhibits an apparent topological representation of the aorta, while achieving a high Dice score with the conventional delineated region (b).

To generate these pseudo labels, we initially propose an efficient, streamlined approach involving several annotation standards to annotate the elliptical structures. Then, we deploy an ellipse-fitting algorithm to obtain the numerical forms of the ellipse-like structures, i.e., the parameters of the ellipses. The ellipses' parameters are then used to generate 2D Gaussian heatmaps, which serve as the pseudo labels (Figure~\ref{fig:s1_1} (d)). The training of DL models with these pseudo labels employs a novel combination of voxel reconstruction loss and distribution loss, supplanting conventional Dice loss \cite{milletari2016v} and binary cross-entropy (BCE) loss. 

This paper contributes by: (1) Introducing a weakly-supervised learning approach that is versatile across different 2D/3D DL models, offering novel insights into ellipse-like VS within non-contrast CTs. The focus in this study centers on abdominal aorta. (2) Proposing a set of annotation standards to reduce labeling time while improving segmentation performance, which not only reduces the need for direct supervision from cardiovascular surgeons but also facilitates the integration of unlabeled public datasets. (3) Presenting an approach for generating pseudo labels, which are then utilized in conjunction with innovative loss functions replacing traditional segmentation loss functions, thereby adapting to the use of these pseudo labels. (4) Exhibiting the superiority of pseudo label in a labeled public dataset.

The remainder of this paper is organized as follows: we present the related works in Section \ref{sec:rw} and elaborate our methodology in Section \ref{sec:method}. The contrastive results of experiments and the qualitative discussion are provided in Section \ref{sec:ex} and Section \ref{sec:dis}, respectively. The Section \ref{sec:con} state the conclusion.

\section{Related Works}\label{sec:rw}
In this section, we survey the literature on conventional DL-based segmentation models, DL-based methods for segmenting the aorta in non-contrast CTs, and current weakly-supervised learning-based methods for segmenting VS.
\subsection{Current DL-based segmentation models}
State-of-the-art DL-based segmentation models mainly contain the pure CNN-based and the CNN-Transformer-hybrid models. Generally, the former is based on an encoder-decoder mechanism constructed by stacked convolutional-normalization-activation layers. The convolution performs in both 2D and 3D dimensions, which yields the origin of 2D and 3D segmentation models, e.g., the U-net \cite{ronneberger2015u}, Attention u-net \cite{oktay2018attention}, Residual U-Net \cite{zhang2018road} (2Ds) and 3D U-net \cite{cciccek20163d}, V-net \cite{milletari2016v} (3Ds). The CNN-Transformer-hybrid models involve Transformer constructs into the conventional CNN segmentors to make it more effective. The Transformer relying on the parallel multi-head attention mechanism \cite{vaswani2017attention} was initially proposed in Natural Language Processing while well performed in medical image processing. The way to integrate the Transformer into CNN is flexible, e.g., using Transformer as a part of the encoder (TransUnet \cite{chen2021transunet}), the whole encoder (Swin UNETR \cite{hatamizadeh2021swin}), or the bottleneck (TransBTS \cite{wang2021transbts}). All the methods above are the current DL-based segmentation models based on strong labels. Besides, the recent rise of the self-adapt segmentation frameworks such as nn-U-net \cite{isensee2021nnu}, AdaResU-Net \cite{baldeon2020adaresu}, outperformed the original models through the self-configuring methods and the data pre-processing. While most of these models have the capability of segmenting anatomical structures, they did not specifically address the issues associated with VS segmentation in non-contrast CTs and weakly supervised scenarios. Consequently, there is an ongoing need to adapt the implementation and evaluate the performance of these DL models in addressing these specific issues.
\subsection{DL-based methods for segmentation of the aorta in non-contrast CTs}
Although aortic segmentation in non-contrast CT scans is a non-trivial problem, there are very few studies based on DL methods in this domain. Lu et al. proposed DeepAAA \cite{lu2019deepaaa}, a derivative of 3D U-net \cite{cciccek20163d}, experimented on a mixture of CTAs (52\%) and non-contrast CTs (48\%) in both training and inference stages, which did not specifically emphasize the case of pure non-contrast CTs. Chandrashekar et al. \cite{chandrashekar2020deep} proposed a 3D cascaded attention-based CNN model that involved additive attention gates while performing as a cascaded way to implement coarse-fine segmentation. They mainly verified the model on their local data containing 26 non-contrast CTs, where a large amount of online data augmentation was applied to achieve considerable results. To address the issues of segmentation of the aorta in non-contrast CTs, in our previous work, we proposed a CNN-based 2D-3D feature fusion mechanism \cite{ma2023deep}, which achieved competitive results involving the presence of aortic aneurysm. All the aforementioned DL-based methods require strong labels for fully-supervised learning, which is labor-intensive and has the potential risk of the effects from intra/inter-observer variability.
\subsection{Weakly-supervised learning and applications in VS}\label{sec:related_pseudo_labels}
Weakly supervised learning generally relies on training with weak labels to achieve competitive performance \cite{zhou2018brief, tajbakhsh2020embracing, tajbakhsh2021guest, ren2023weakly}. According to Nima et al., \cite{tajbakhsh2020embracing}, weak labels can be roughly categorized into the following types: firstly, image-level labels, represented by Class activation maps (CAMs), generate salient maps through the feature maps of pre-trained models \cite{zhou2016learning, selvaraju2017grad}. The boundaries of these maps are then determined through expansion or restriction \cite{ahn2019weakly, wei2017object}. The second type is sparse labels, such as bounding boxes \cite{khoreva2017simple}, scribbles \cite{lin2016scribblesup}, and sparse points \cite{matuszewski2018minimal}. These sparse annotations are generally smaller subsets of their corresponding strong labels, consuming less labeling time. Recent research on weakly supervised medical image segmentation has emphasized the use of scribbles as annotations. By densely combining \cite{wang2023weakly}, human interventions in difficult areas \cite{zhuang2024annotation}, the superpixel-guided scribble walking \cite{zhou2023weakly} effectively leverages pseudo labels to enhance segmentation performance. The third type is noisy labels, which retain the general structure of strong labels but lack explicit boundaries \cite{gu2018reliable}. These labels undergo refinement \cite{min2019two} or are directly used in training in combination with robust loss functions \cite{mirikharaji2019learning}.

In recent years, weakly supervised methods for vascular structure segmentation have gained increasing attention. Some of these methods focus on 2D images, such as vessel segmentation in 2D X-ray cerebral \cite{vepa2022weakly} and coronary angiography \cite{zhang2020weakly}, utilizing active contour models and uncertainty estimation.  \cite{guo20243d} employed 2D projection annotations to generate 3D pseudo labels, applying this approach to weakly supervised learning in aortic CTA. Some studies have leveraged intrinsic image characteristics to provide prior knowledge for weak supervision, such as vessel segmentation in 2D Doppler images \cite{ning2023doppler} and 2D laser speckle contrast images \cite{fu2023robust}, with the former targeting the radial and carotid arteries and the latter focusing on rabbit ear blood vessels. Additionally, some research has exploited the complex tree-like characteristics of vascular structures. For instance, \cite{wu2022weakly} enhanced model learning of tubular structures by incorporating Hessian shape priors, facilitating 3D cerebrovascular segmentation. \cite{xu2023extremely} used manually labeled small tree structures and generative models to create synthetic kidney vascular trees, aiding in the segmentation of vessels in 3D micro-CT scans of rat kidneys. \cite{zhu2024tsp} proposed a metric based on vascular tree topology and demonstrated its effectiveness in weakly supervised learning on hepatic vessels in the liver-hepatic CT dataset. These approaches highlight the potential of weakly supervised learning for vascular structure segmentation. However, to the best of our knowledge, no prior work has extended weakly supervised learning to the segmentation of non-contrast-enhanced aortic structures.
\section{Method} \label{sec:method}
\subsection{Overall Framework}
\begin{figure*}[!t]
\centering
% \begin{center}{
\begin{minipage}[c]{0.85\linewidth}
\includegraphics[width=\textwidth]{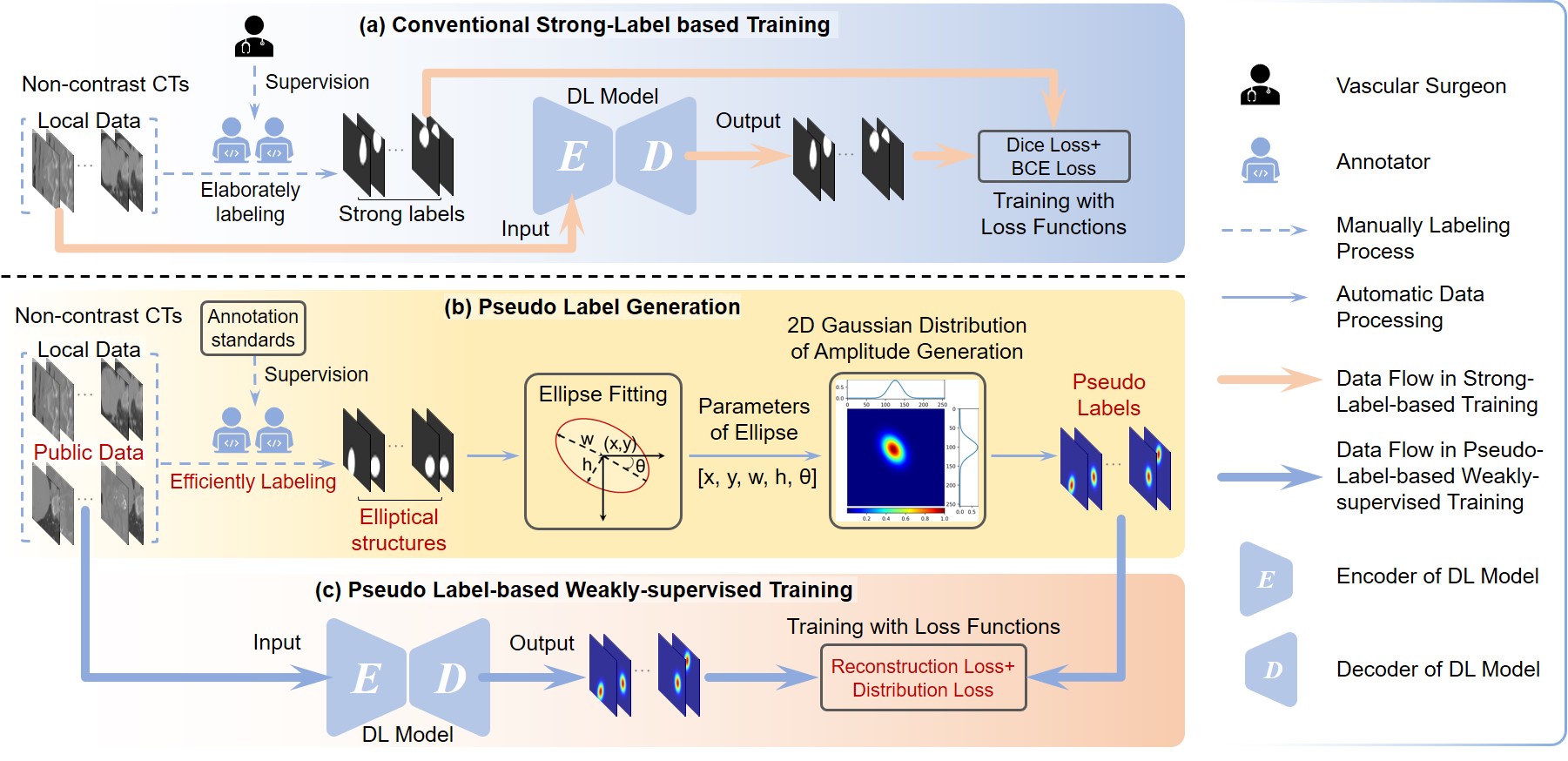}
\end{minipage}
% }\end{center}
\caption{Comparison of (a) strong-label-based training and (b)-(c) proposed pseudo-label-based weakly-supervised training approaches. In (a), conventional strong-label-based training requires time-intensive, expert-elaborated labeling and heavy supervision of vascular surgeons for strong labels, employing Dice and BCE loss for model optimization. For the proposed method, (b) shows the generation of pseudo label. Based on the proposed annotation standards, elliptical structures are efficiently annotated by the experts. The elliptical structures are then processed via an ellipse-fitting algorithm to establish five foundational parameters: the location of the central point $(x,y)$, the semi-major and semi-minor axes $w$ and $h$, and the rotation angle $\theta$. These parameters create 2D Gaussian heatmaps with pixel intensities in $[0,1]$, employing a constant to restrict intensities exceeding $0.5$ within the ellipse boundary. The Gaussian heatmaps serve as pseudo labels, generated through a weak but efficient process. With the annotation standards and efficiency, external public data can be incorporated to enrich the training set without exhaustive annotator efforts or heavy supervision of surgeons. Consequently, we regard the subsequent training in (c) as a weakly-supervised training strategy. It adopts a novel combination of a voxel reconstruction loss and a distribution loss to adapt the pseudo labels for model optimization.}\label{fig:s3_1}
\end{figure*}

We illustrate the distinction between conventional strong-label-based training and proposed pseudo-label-based weakly-supervised training pipelines in Figure~\ref{fig:s3_1}. The conventional strong-label-based training consumes a large expenditure of time and effort for the annotators (experts) to make elaborately annotate the non-contrast CTs, demanding heavy supervision by vascular surgeons due to its time and labor intensity. This method is susceptible to generate intra/inter-observer variability because of the inherent ambiguity of boundaries in non-contrast CTs. The strong labels are binary masks, employing Dice loss \cite{milletari2016v} and binary cross-entropy (BCE) loss to train and optimize the DL models. Conversely, the proposed weakly-supervised method entails annotators delineating elliptical structures based on our proposed annotation standards, substantially reducing labeling time and the reliance of supervision from vascular surgeons. This method accommodates external (public) data through the proposed annotation standards and efficient labeling. The delineated elliptical structures are then processed by an ellipse-fitting algorithm to yield foundational parameters, utilized to generate Gaussian heatmaps as pseudo labels. During training, a novel combination of voxel reconstruction and distribution losses optimize DL models using these pseudo labels. The following parts elaborate the proposed pseudo-label-based weakly-supervised training in terms of pseudo label generation and pseudo-label-based weakly-supervised training.
\subsection{Pseudo Label Generation} \label{subsec:p_l_g}
As Figure~\ref{fig:s3_3} illustrates, the generation of pseudo label includes 1) efficiently labeling with annotation standards, 2) ellipse fitting and 3) Gaussian heatmap generation, where the first step is the process with manual work while the last two steps are fully automatic approaches implemented by computer-assisted image processing techniques. Therefore, reducing manually labeling time in the first steps is crucial for the efficiency of pseudo label generations.
\subsubsection{Efficiently Labeling with Annotation Standards}\label{subsec:as}
\begin{figure}[!t]
% \begin{center}{
\begin{minipage}[c]{1.0\linewidth}
\includegraphics[width=\textwidth]{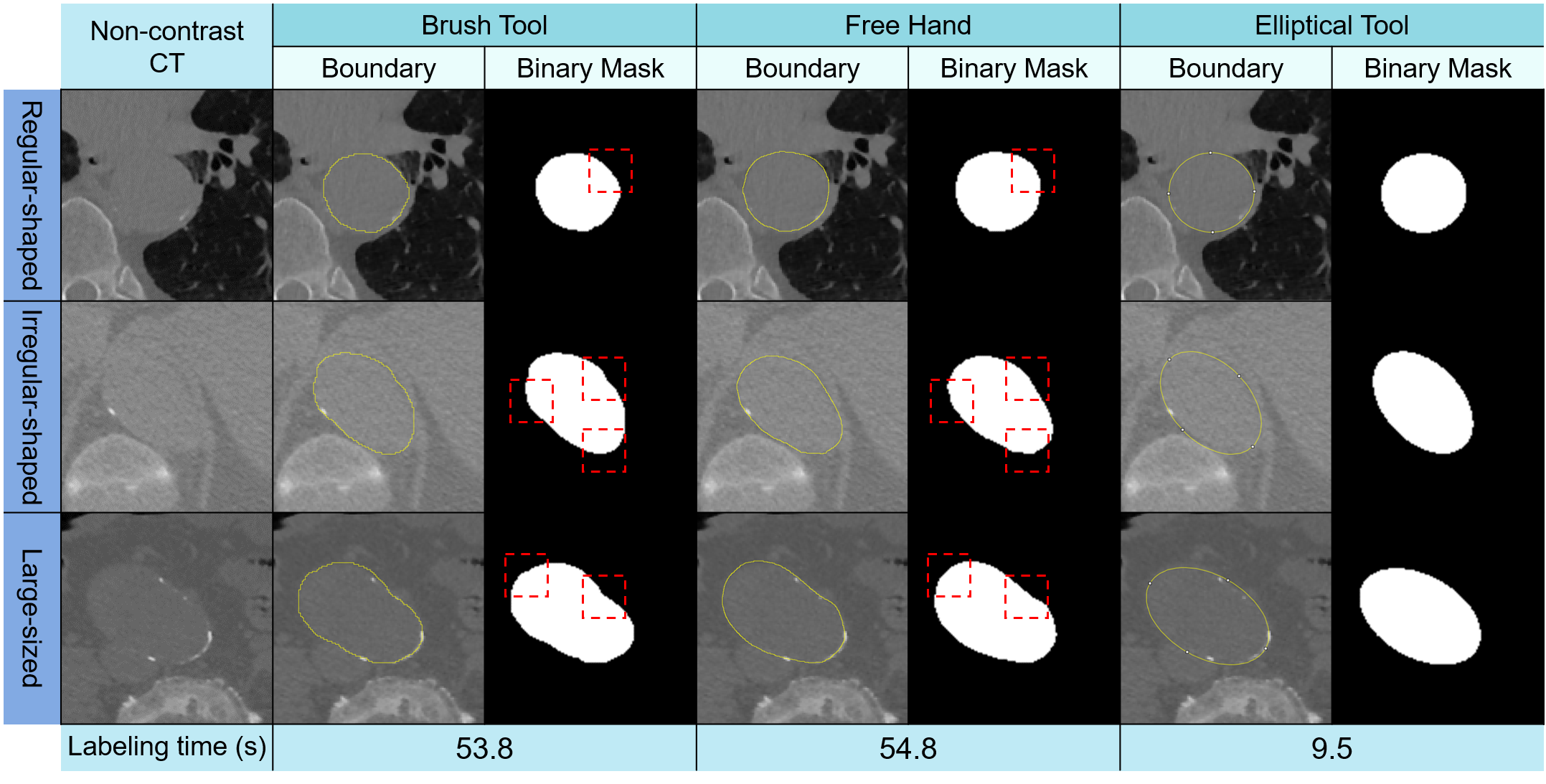}
\end{minipage}
% }\end{center}
\caption{Annotations by a single expert using ImageJ \cite{schneider2012nih} displayed in boundary and binary masks. The samples are randomly selected from three types of abdominal aortas in non-contrast CT slices: regular-shaped (circular), irregular-shaped (elliptical), and large-sized (aneurysm-contained) aortas. The annotations are obtained from three mechanisms of ImageJ: 1) \textbf{Brush Tool} uses a tiny draggable circular brush for target region filling, and 2) \textbf{Free Hand} delineates along boundaries, the red dotted boxed indicates intra-observer variability of the two approaches, and 3) \textbf{Elliptical Tool}, employed in our approach, selects elliptical regions. The series of intra-observer variability between Brush Tool and Elliptical Tool are marked by the red dotted box. Annotations via Elliptical Tool adhere to the proposed annotation standards, depicting stable topologies. The average labeling time per slice of each tool is showed in last row.}\label{fig:s3_2}
\end{figure}
In pursuit of efficient annotation, we introduce several annotation standards to guide annotators in delineating ellipse structures in non-contrast CT images, where we specify that \textbf{selected areas} are annotated regions and \textbf{target areas} are the ideal regions representing the VS in the CT slices.

1) \textbf{Closed Conic Section Annotation.} This entails outlining selected areas as closed conic sections, i.e., ellipses or circles. These closed conic sections accurately represent aortas' topologies in CT slices. Annotation toolkits typically offer Circle and Ellipse annotation tools, enabling annotators to efficiently mark target regions.

2) \textbf{Complete Coverage.} It ensures that the selected area completely covers the target area, maintaining its topological integrity. Moreover, it addresses the ambiguity of aorta boundaries in non-contrast CTs. The indistinct boundary of the aorta impedes the annotation in non-contrast CTs, while comprehensive coverage mitigates it by encompassing the extension of the unclear boundary. It relieves the burden on annotators by reducing boundary judgment while decreasing the intra/inter-observer variability.

3) \textbf{Minimum Perceptible Difference.} Based on the two conditions mentioned above, this principle focuses on retaining the least perceptible difference between the selected and target areas to mitigate false positives.

Following these annotation standards, the best selected area is assumed to be the minimal external ellipse of the corresponding target area.

We use ImageJ \cite{schneider2012nih}, a lightweight public image labeling toolkit as an instance. Figure~\ref{fig:s3_2} illustrates the annotations delineated by a single expert using three types of labeling mechanisms provided by ImageJ, where the Brush Tool and Free Hand aim for delineating strong labels and the Elliptical Tool is for elliptical structures. We adopt the Elliptical Tool as an annotation process of our method, of which the samples are generated following the three proposed annotation standards. According to Figure~\ref{fig:s3_2}, the strong labels obtained by Brush Tool and Free Hand inevitably exhibit the intra-observer variability due to the ambiguous boundaries, while the ones from elliptical Tool manifest stable topologies. The average labeling time for each slice of Elliptical Tool is $9.5$s, $82.3\%$ decreased compared to the other two approaches. Moreover, we observe that neither the Brush Tool nor Free Hand can generate an annotation by a one-shot delineation in our practice, where each annotation requires a series of refinements. It consequently requires more supervision of surgeons and domain knowledge for the judgment of the boundaries while consuming more labeling time.
\subsubsection{Ellipse Fitting}
The obtained annotation is visually an ellipse-like binary mask and formally a set of data points. To further generate a pseudo label possessing the characteristic of an ellipse, the numerical form of the ellipse needs to be determined. Therefore, as it shows in Figure~\ref{fig:s3_3} (b), the ellipse-like binary mask is used to evaluate the parameters which define a correspondent numerical ellipse, i.e., central point $(x,y)$, the major and minor axes $w$ and $h$, and the rotation angle $\theta$.

An ellipse is a particular case of conic curve which can be numerically formulated as a second-order polynomial
\begin{equation}
{F_{\bf{a}}}({\bf{x}}) = {\bf{x}} \cdot {\bf{a}} = 0,
\end{equation}
where ${\bf{a}} = {[a,b,c,d,e,f]^{\rm T}}$ and ${\bf{x}} = [{x^2},xy,{y^2},x,y,1]$, with a constraint specifically for ellipse
\begin{equation}\label{equ:cons1}
{b^2} - 4ac < 0,
\end{equation}
the $a$,$b$,$c$,$d$,$e$, and $f$ are coefficients of the ellipse, and $(x,y)$ are the coordinates of data points that lie on it. In our case, the points $(x,y)$ are the coordinates of the boundaries of the obtained ellipse-like regions. 

We leverage the method proposed by \cite{haralick1992computer} to fit a general conic to the set of points $(x_i,y_i)$, $i=1...N$ by minimizing the sum of the squared distances of the points to the conic defined by $\bf{a}$. Then the process is simplified and stabilized by the improved least squares method of \cite{halir1998numerically}. We finally obtain ${\bf{a}} = [a,b,c,d,e,f]^{\rm T}$ represents the coefficients of the best-fit numerical ellipse for the given data points. As a result, the five parameters of the central point $(x,y)$, the semi-major and semi-minor axes $w$ and $h$, and the rotation angle $\theta$ of the fitted ellipse can be determined by the coefficient vector $\bf{a}$ \cite{weisstein2014ellipse}.

\subsubsection{Gaussian Heatmap Generation}
With the five fundamental elliptical parameters, the heatmaps of 2D Gaussian can be generated as pseudo labels.We elaborately assign the heatmap with its intensity ${\bf{{\rm I}}} \in [0,1]$, where the pixels that are enclosed by the ellipse curve contain intensities larger than 0.5 while the outliers are less than 0.5. This mechanism corresponds to the binary segmentation, with the activation of the sigmoid function mapping the output of the model into $[0,1]$, the regions containing intensities ${\bf{{\rm I}}} > 0.5$ are regarded as the predicted results. Generally, the Bivariate Gaussian Probability Density Function (PDF) is expressed as
\begin{equation}\label{equ:gau1}
f({\bf{X}}) = \frac{1}{{\sqrt {2\pi \left| \bm{\Sigma}  \right|} }} \times {e^{ - \frac{1}{2}{{({\bf{X}} - {\bm{\mu}})}^{\rm T}}{\bm{\Sigma} ^{ - 1}}({\bf{X}} - {\bm{\mu }})}},
\end{equation}
where ${\bf{X}} = {[x,y]^{\rm T}} \sim \mathcal{N}({\bm{\mu }},\bm{\Sigma} )$ contains two random variables in two orthogonal dimensions. ${\bm{\mu }} \in {\mathbb{R}^2}$ is the mean vector defining the location of the central point. $\bm{\Sigma} \in {\mathbb{R}^{2 \times 2}}$ is a positive semi-definite matrix representing the covariance matrix of the two variables. As a real symmetric matrix, $\bm{\Sigma}$ can be orthogonally diagonalized as
\begin{equation}
\bf{\Sigma}  = {\bf{Q}}\Lambda {{\bf{Q}}^{\rm T}} = ({\bf{Q}}{\Lambda ^{{1 \mathord{\left/
 {\vphantom {1 2}} \right.
 \kern-\nulldelimiterspace} 2}}}){({\bf{Q}}{\Lambda ^{{1 \mathord{\left/
 {\vphantom {1 2}} \right.
 \kern-\nulldelimiterspace} 2}}})^{\rm T}},
\end{equation}
where $\bf{Q}$ is a real orthogonal matrix, and $\bm{\Lambda}$ is a diagonal matrix containing the eigenvalues of descending order. The Gaussian probability density function, therefore, can be reformulated as
\begin{equation}
f({\bf{X}}) = \frac{1}{{\sqrt {2\pi \left| {{\bf{Q}}\bm{\Lambda} {{\bf{Q}}^{\rm T}}} \right|} }} \times {e^{ - \frac{1}{2}{{[{{({\bf{Q}}{\bm{\Lambda} ^{{1 \mathord{\left/
 {\vphantom {1 2}} \right.
 \kern-\nulldelimiterspace} 2}}})}^{\rm T}}({\bf{X}} - {\bm{\mu }})]}^{\rm T}}[{{({\bf{Q}}{\bm{\Lambda} ^{{1 \mathord{\left/
 {\vphantom {1 2}} \right.
 \kern-\nulldelimiterspace} 2}}})}^{\rm T}}({\bf{X}} - {\bm{\mu }})]}}.
\end{equation}

The mean vector $\bm{\mu }$, orthogonal matrix $\bf{Q}$, and diagonal matrix $\bm{\Lambda}$ are spatially determined by the ellipse's five parameters evaluated by the ellipse fitting algorithm, where $\bm{\mu } = {[x,y]^{\rm T}}$ represents the central location. The $\bf{Q}$ is a rotation matrix defined by the rotation angle $\theta  \in [ - \frac{\pi }{2},\frac{\pi }{2}]$. The diagonal matrix $\bm{\Lambda}$ contains the eigenvalues $\lambda _1 = w^2$, $\lambda _2 = h^2$,  corresponding to the scale of the ellipse.
\begin{figure*}[!t]
\centering
\begin{minipage}[c]{0.7\linewidth}
\includegraphics[width=\textwidth]{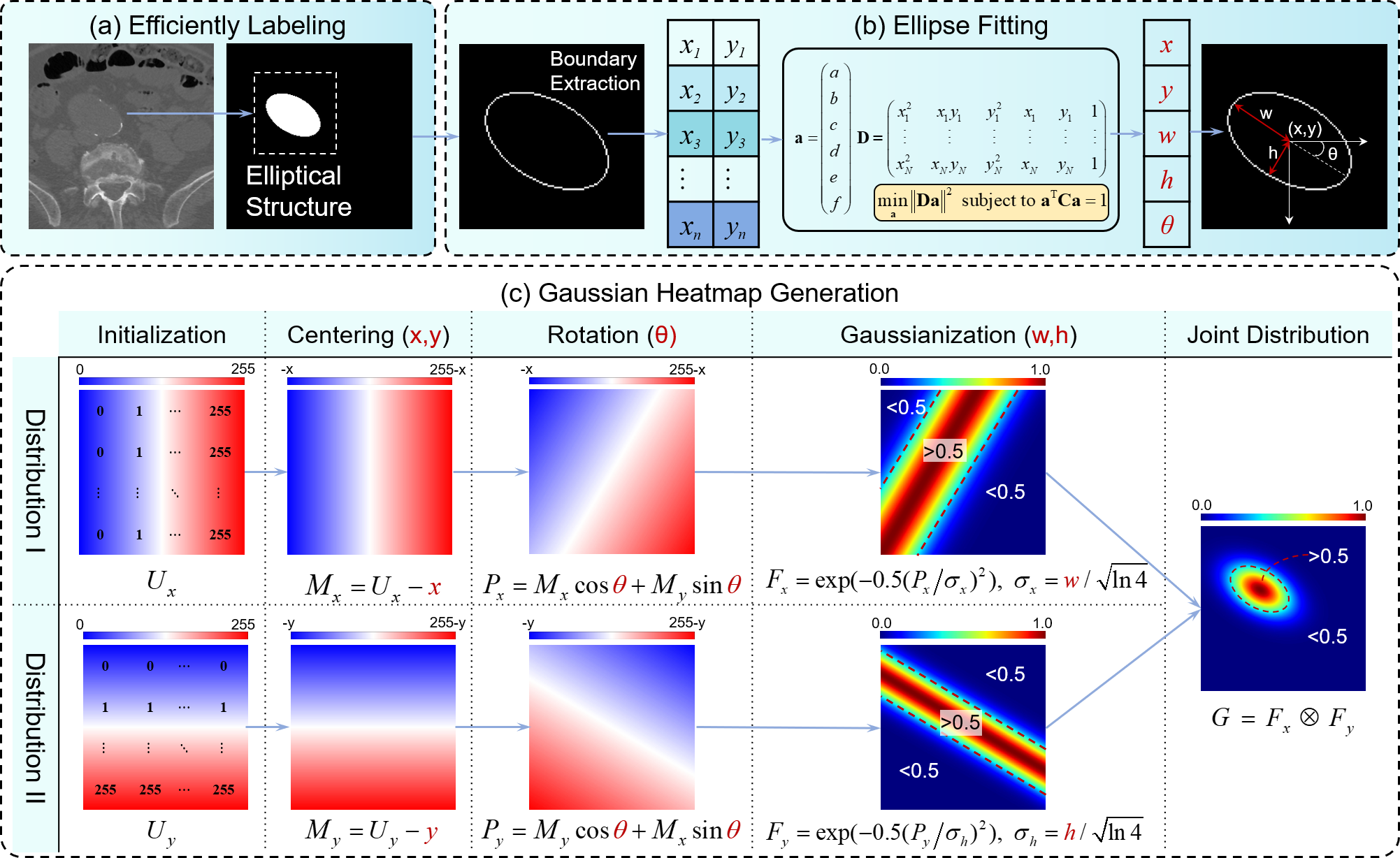}
\end{minipage}
\caption{Process of pseudo label generation, including (a) the efficient labeling of ellipse-like structures based on the proposed annotation standards, (b) ellipse-fitting to obtain the five parameters numerically defining the ellipse, and (c) 2D Gaussian heatmap generation based on the elliptical parameters. The generated Gaussian heatmap contains the pixel intensities of $[0,1]$, used as the pseudo label for the weakly-supervised learning. Note that only step (a) is manually performed, while steps (b) and (c) are fully automatic processes.}\label{fig:s3_3}
\end{figure*}

As Figure~\ref{fig:s3_3} (c) shows, based on the above-mentioned theoretical foundations and the five parameters of the ellipse ($x,y,w,h,\theta$), a 2D Gaussian heatmap with specific distribution of pixel intensities can be generated by the following steps. 

1)  Initialization - Initializing two discrete uniform distributions in two orthogonal directions, respectively. Assuming that the spatial size is $256\times 256$, the two matrix of discrete uniform distributions $U_x$ and $U_y$ are
% \begin{equation}
% {U_x} = \left( {\begin{array}{*{20}{c}}
% {0}&{1}& {\cdots}& {255}\\
% {0}&{1}& {\cdots}& {255}\\
%  {\vdots}&  {\vdots}&  {\ddots}&  {\vdots} \\
% {0}&{1}& {\cdots}& {255}
% \end{array}} \right){\rm{, }}{U_y} = \left( {\begin{array}{*{20}{c}}
% 0&0& \cdots& 0\\
% 1&1& \cdots& 1\\
%  \vdots&  \vdots&  \ddots&  \vdots \\
% {255}&{255}&{255}&{255}
% \end{array}} \right).
% \end{equation}
\begin{equation}
{U_x} = \left( {\begin{array}{*{20}{c}}
{0}&{\cdots}& {255}\\
 {\vdots}&{\ddots}&{\vdots}\\
{0}&{\cdots}& {255}
\end{array}} \right){\rm{, }}{U_y} = \left( {\begin{array}{*{20}{c}}
0&\cdots& 0\\
{\vdots}&{\ddots}&{\vdots}\\
{255}&{\cdots}&{255}
\end{array}} \right).
\end{equation}

2) Centering - Localizing the central position by subtracting the coordinates of the central point $(x,y)$
\begin{equation}
{M_x} = {U_x} - x{\mbox{, }}{M_y} = {U_y} - y.
\end{equation}
 
3) Rotation - Rotating the distributions with the rotation angle $\theta$
\begin{equation}
{P_x} = {M_x}\cos \theta  + {M_y}\sin \theta {\mbox{, }}{P_y} = {M_y}\cos \theta  + {M_x}\sin \theta.
\end{equation}

4) Gaussianization - Generating two Gaussian distributions with the semi-major/minor axis $w$ and $h$, respectively. The two Gaussian heatmaps contain pixel intensities in $[0,1]$, where the intensities exceeds 0.5 if the pixels are inside the region of $[{P_x}-w, {P_x}+w]$, $[{P_y}-h, {P_y}+h]$ in two heatmaps, respectively.

We initially define the Gaussian Probability Density Functions (PDFs) of the two orthogonal directions
\begin{equation}
{f_{Px}}(t) = \frac{1}{{{\sigma _x}\sqrt {2\pi } }}{e^{ - \frac{{{{(t - x)}^2}}}{{2\sigma _x^2}}}}{\mbox{, }}{f_{Py}}(t) = \frac{1}{{{\sigma _y}\sqrt {2\pi } }}{e^{ - \frac{{{{(t - y)}^2}}}{{2\sigma _y^2}}}},
\end{equation}
then, we perform 0-1 normalization for the two PDFs. Since the peak values are not zero, we obtain
\begin{equation}\label{equ:norm}
{F_x} = \frac{{{f_{Px}}(t)}}{{{f_{Px}}(t = x)}} = {e^{ - \frac{{{{(t - x)}^2}}}{{2\sigma _x^2}}}}{\mbox{, }}{F_y} = \frac{{{f_{Py}}(t)}}{{{f_{Py}}(t = x)}} = {e^{ - \frac{{{{(t - y)}^2}}}{{2\sigma _y^2}}}}.
\end{equation}
To control the boundary of threshold $0.5$, let $t-x=w$, $t-y=h$, $F_x=F_y=0.5$, it is easy to get
\begin{equation}
{\sigma _x} = \frac{w}{{\sqrt {\ln 4} }}{\mbox{, }}{\sigma _h} = \frac{h}{{\sqrt {\ln 4} }}.
\end{equation}
Let $t-x=P_x$, $t-y=P_y$, the two orthogonal Gaussian heatmaps $F_x$ and $F_y$ of Equ.(\ref{equ:norm}) can be consequently expressed as
\begin{equation}
{F_x} = {e^{ - \frac{{P_x^2}}{{2{{({w \mathord{\left/
 {\vphantom {w {\sqrt {\ln 4} }}} \right.
 \kern-\nulldelimiterspace} {\sqrt {\ln 4} }})}^2}}}}} = {e^{ - \ln 2\frac{{P_x^2}}{{{w^2}}}}},{F_y} = {e^{ - \frac{{P_y^2}}{{2{{({h \mathord{\left/
 {\vphantom {h {\sqrt {\ln 4} }}} \right.
 \kern-\nulldelimiterspace} {\sqrt {\ln 4} }})}^2}}}}} = {e^{ - \ln 2\frac{{P_y^2}}{{{h^2}}}}}.
\end{equation}

5) Integration - Integrating the two orthogonal distributions into a joint distribution to obtain the 2D Gaussian elliptical heatmap $G$
\begin{equation}
G = {F_x} \otimes {F_y},
\end{equation}
where $\otimes$ is the element-wise multiplication of the pixels. We use the generated 2D Gaussian elliptical heatmap as the Pseudo label for the weakly-supervised learning of DL models.
\subsection{Pseudo-label-based Weakly-supervised Training}\label{sec:loss}
The generated pseudo labels aim to be used to train various 2D/3D DL models. Since the generation of the pseudo labels is a 'weakly' process with efficient annotation, we regard our learning strategy as weakly-supervised learning. Different from the binary mask of strong label, our pseudo label contains a 2D Gaussian distribution while each pixel intensity represents a normalized probability density, where a pixel possessing a value larger than $0.5$ means that it has a large possibility to be the foreground. Therefore, an ideal pattern learned by a model should be a 2D Gaussian distribution where each pixel contains a specific normalized probability density. To fit both the distribution and the probability density, we involve a novel combination of loss functions to supplant conventional Dice loss and BCE loss for segmentation.

1) \textbf{Distribution loss.} We initially involve the Kullback–Leibler (KL) divergence \cite{kullback1951information} loss to fit the 2D Gaussian distribution of the pseudo label in each slice. Let ${\rm X}^{n \times n}$, $G^{n \times n}$ be the 2D output of a DL model and the corresponding pseudo label, respectively. The KL divergence can be expressed as
\begin{equation}
{\mathcal{L}_{KL}}({P_G}\parallel {P_X}) = {P_G}\log ({{{P_G}} \mathord{\left/
 {\vphantom {{{P_G}} {{P_X}}}} \right.
 \kern-\nulldelimiterspace} {{P_X}}}),
\end{equation}
where $P_G$ is the 2D Gaussian distribution of pseudo label $G$ and $P_X$ is the distribution of the output $X$. In this case, the distributions $P_G$ and $P_X$ can be obtained as maps of probability through the Softmax function. The probability of the pixel at the position $(i,j)$ is
\begin{equation}
{P_{{G_{(i,j)}}}} = \sigma ({G_{(i,j)}}) = {{{e^{{G_{(i,j)}}}}} \mathord{\left/
 {\vphantom {{{e^{{G_{(i,j)}}}}} {\sum\limits_{x = 1}^n {\sum\limits_{y = 1}^n {{e^{{G_{(x,y)}}}}} } }}} \right.
 \kern-\nulldelimiterspace} {\sum\limits_{x = 1}^n {\sum\limits_{y = 1}^n {{e^{{G_{(x,y)}}}}} } }},
\end{equation}
\begin{equation}
{P_{{X_{(i,j)}}}} = \sigma ({X_{(i,j)}}) = {{{e^{{X_{(i,j)}}}}} \mathord{\left/
 {\vphantom {{{e^{{X_{(i,j)}}}}} {\sum\limits_{x = 1}^n {\sum\limits_{y = 1}^n {{e^{{X_{(x,y)}}}}} } }}} \right.
 \kern-\nulldelimiterspace} {\sum\limits_{x = 1}^n {\sum\limits_{y = 1}^n {{e^{{X_{(x,y)}}}}} } }}.
\end{equation}

For the cases of 3D models, instead of fitting the data as slice-by-slice 2D Gaussian distribution, we regard the data to be fit as a high dimensional distribution within a 3D volume. Let ${\rm X}^{d\times n \times n}$, $G^{d\times n \times n}$ be the 3D output the and the 3D pseudo label stacked by the 2D ones, respectively. We leverage the Wasserstein loss to deal with this case
\begin{equation}
{\mathcal{L}_{Wa}} = {\inf _{\gamma \in \prod {(P_x,P_G)} }}{\mathbb{E}_{(x,y) \sim \gamma }}\left[ {\left\| {x - y} \right\|} \right],
\end{equation}
where ${\gamma \in \prod {(P_X,P_G)} }$ represents the set of all joint distributions $\gamma (x,y)$ whose marginal distributions are $P_G$ and $P_X$. This loss function indicates the minimization of the Expectation $\mathbb{E}$ of $\gamma (x,y)$ to move $x$ to $y$ in order to transform the distribution $P_X$ into the distribution $P_G$.

2) \textbf{Reconstruction loss.} To fit each pixel/voxel possessing the probability density, we use Mean Absolute Error (MAE) loss to make the reconstruction
\begin{equation}
{\mathcal{L}_{Rec}} = \frac{1}{t}\sum\limits_{i = 1}^t {\left| {G - X} \right|},
\end{equation}
where $t$ is the number of pixels/voxels of $G$ and $X$ in 2D/3D cases, respectively.

The overall loss is the combination of the two losses with weights
\begin{equation}\label{equ:loss_all}
\mathcal{L} = {w_1}{\mathcal{L}_{Dis}} + {w_1}{\mathcal{L}_{{\mathop{\rm Re}\nolimits} c}},
\end{equation}
where $\mathcal{L}_{Dis}$ is $\mathcal{L}_{KL}$ and ${\mathcal{L}_{Wa}}$ in 2D and 3D cases, respectively.

% \subsection{Option: Fine-tuning for Downstream Tasks}
\section{Experiments} \label{sec:ex}
\subsection{datasets}
There are three datasets involved in this study, i.e., the local data set, the public data sets Medical Segmentation Decathlon (MSD) and TotalSegmentator. We regard the local set and the MSD as label-agnostic datasets while TotalSegmentator as the label-provided dataset. For the experiments on label-agnostic datasets, we regard the MSD as an external dataset of the local data.
\subsection{Local dataset containing AAAs and external dataset}
The first dataset is our local data collected retrospectively at Rennes University Hospital from patients who underwent the EVAR procedure. Patient's informed consent was obtained for anonymous registration in the research database. The local data was obtained from 30 patients suffering from abdominal aortic aneurysms (AAAs), where a pre-operative non-contrast-enhanced CT scan was performed on each patient. The original imaging data were given in Digital Imaging and Communications in Medicine (DICOM) format, containing a spatial size of 512 × 512 and a thickness of 0.625 to 5 mm for each axial slice.

Two experts ($A$ and $B$) generated the pseudo labels and manually delineated the strong labels of the local dataset. Expert $A$ delineated all the non-contrast CTs, obtaining the pseudo labels $p$ and strong labels $s$, respectively. Note that the pseudo labels $p$ were generated through the proposed annotation standards, ellipse fitting, and Gaussian heatmap generation while the delineation of strong labels $s$ was supervised by a vascular surgeon. To evaluate the intra- and inter-observer variability of the manual segmentation, following the related work \cite{chandrashekar2020deep}, we randomly selected a subset $t$ of the local data ($|t|=10$). The expert $B$ annotated $t$ independently, generating the pseudo label $p_B$ and strong label $s_B$. The expert $A$ annotated $t$ for a second time after a gap of 10 days to generate the $p_A$ and $s_A$. The aforementioned annotation process was implemented using ImageJ \cite{schneider2012nih}. As presented in section \ref{subsec:p_l_g}, the pseudo labeling involved the Elliptical Tool, while the strong labels were delineated through a combination of Free Hand and Brush Tool.

Consequently, the $p_A/s_A$ and $p_B/s_B$ were compared against $p/s$ in terms of Dice score to assess the intra-/inter-observer variability of the manual annotation, respectively. Table~\ref{tab:intra_inter} shows the intra-/inter-observer variability of both pseudo labels $p$ and strong labels $s$. It manifests that pseudo labels contain higher intra-/inter-observer variability, supporting its reliability and stability for DL-based models' training.
\begin{table}[h]
\caption{Intra and inter-observer variability of pseudo labels $p$ and strong labels $s$, in terms of Dice score (\%).}
   \begin{center}
  \begin{tabular}{ccc}
  \hline
  Labels      &Intra-   &Inter-\\
  \hline
   Pseudo $p$ &97.8$\pm$1.1&97.0$\pm$1.3\\
  \hline
   Strong $s$ &96.6$\pm$1.1&96.1$\pm$1.4\\
\hline
  \end{tabular}
  \end{center}
  \label{tab:intra_inter}
\end{table}

A 256$\times$256 Region of Interest (RoI) of uniform spatial position was extracted automatically by center-cropping from each slice to improve the training and inference efficiency. We obtained 30 volumes of local data containing 5749 axial slices, accompanied by pseudo labels $p$ and strong labels $s$. The volumes were divided into three subsets (non-overlapping for patients), marked as $D_0$, $D_1$, and $D_2$ for 3-fold cross-validation. Note that the pseudo labels are only used for training, while the strong labels are used for validation and testing.

\begin{figure}[!t]
% \begin{center}{
\begin{minipage}[c]{1.0\linewidth}
\includegraphics[width=\textwidth]{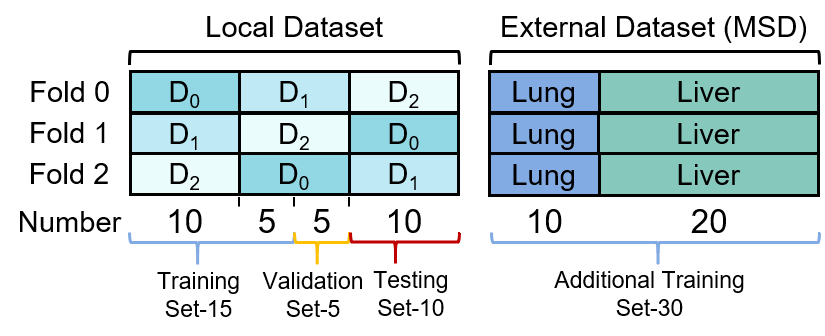}
\end{minipage}
% }\end{center}
\caption{The division of dataset for 3-fold cross-validation. In each fold, the local data are separated to 15 samples (volumes) for training, 5 for validation and the rest 10 for testing. There are 30 samples of external data MSD, including 10 Lungs and 20 Livers, serving as additional training set. Note that all the training and additional training sets are trained with pseudo labels while the inference in validation and testing sets are evaluated by strong labels.}\label{fig:s4_1}
\end{figure}

Based on the proposed annotation standards and the efficiency of generating pseudo labels, we assume that introducing external data to enrich the training set will not cost exhaustive annotator efforts but will improve the performance of DL models. Therefore, we involved the public data Medical Segmentation Decathlon (MSD) \cite{antonelli2022medical} to serve as the additional training set. We chose two subsets of MSD, i.e., Liver and Lung, because the abdominal aortas are well exhibited in these two subsets. To keep the balance of external and local data, we randomly chose 30 samples (10 Lungs and 20 Livers) of MSD to be the additional training set. The same pre-processing as local data was performed on it. Note that we only generated pseudo labels for the external dataset MSD, and there are no AAAs exhibiting in this dataset. We present the data size and labeling time of local and external data in Table~\ref{tab:data_num}. The division for 3-fold cross-validation is showed in Figure~\ref{fig:s4_1}.
\begin{table}[h]
\setlength{\tabcolsep}{0.9mm}
\caption{Data size and labeling time of pseudo labels ($p$) and strong labels ($s$) of local dataset, Medical Segmentation Decathlon (MSD), and TotalSegmentator (TS).}
   \begin{center}
  \begin{tabular}{cccccc}
  \hline
  Dataset  &Volumes   &Slices &\makecell{Label\\type} &\makecell{Public strong\\labels offered} &\makecell{Labeling\\time (h)}\\
  \hline
   \multirow{2}{*}{Local} &\multirow{2}{*}{30}&\multirow{2}{*}{5749}&$s$&~&84.6\\
   ~ &~&~                                                           &$p$&~&15.2\\
  \hline
   MSD &30&5064                                                     &$p$&~&13.3\\
   \hline
   \multirow{2}{*}{TS} &\multirow{2}{*}{60}&\multirow{2}{*}{6944}&$s$&\multirow{2}{*}{\checkmark}&/\\
   ~ &~&~                                                        &$p$&~                          &/\\
\hline
  \end{tabular}
  \end{center}
  \label{tab:data_num}
\end{table}
\subsection{Public dataset TotalSegmentator}\label{subsec:totalseg}
In order to validate the generalization of the proposed method, we also employed the public dataset TotalSegmentator \cite{wasserthal2023totalsegmentator}, originally created by the department of Research and Analysis at University Hospital Basel. The raw data encompasses 117 categories. We randomly sampled 60 volumes, specifically selecting the portions containing the abdominal aorta, with a RoI at 256$\times$256 and number of slices ranging between 65 to 184. The voxel spacing is unified as 1.5 mm in each direction.

Among the sampled volumes, 30 were randomly assigned as the training set, 10 as the validation set, and the remaining 20 as the test set.  It is noteworthy that TotalSegmentator provides strong labels for the aorta. Leveraging these labels, we directly conducted ellipse fitting and generated Gaussian heatmaps to create pseudo labels (step (b) and (c) of Figure~\ref{fig:s3_3}). In this case, the pseudo label generation process can be considered fully-automatic, as it does not incur any additional manual annotation time, given the availability of strong labels in TotalSegmentator. The data size is shown in Table~\ref{tab:data_num}.
\subsection{Implementation Details and Optimization}
We apply the pseudo-label-based weakly-supervised learning in several DL models. To assess the general applicability, the involved models encompass both CNN- or Transformer-based 2D/3D architectures, including Attention U-Net \cite{oktay2018attention}, TransUNet \cite{chen2021transunet}, SwinUNETR \cite{hatamizadeh2021swin}, 3D U-net \cite{cciccek20163d}, and TransBTS \cite{wang2021transbts}. The models are trained for an epoch of $n=500$ and stopped by an early-stopping strategy. The weights that generate the minimum loss in the validation set are used for the inference in the test set. The adam optimization \cite{kingma2014adam} is employed with an initial learning rate of $r=0.001$, linearly decreasing by a ratio of $1-r \times n$ for each epoch, with $\beta_1=0.9$, $\beta_2=0.999$, weight decay is $1\times 10^{-10}$ during training. The weights of loss functions Eq.(\ref{equ:loss_all}) are set to $w_1=w_2=1$ for all the experiments, implemented by Pytorch \cite{paszke2019pytorch}, deployed on Ubuntu 20.04 with a GPU of Nvidia GeForce GTX 1080 (12 GB memory). The input size is $256\times 256$ with a batch size $N=16$ for 2D models and $16\times 256\times 256$ (50\% overlap between successive inputs) with a batch size $N=1$ for 3D models. Before input, the input slice/volume intensities are truncated in the 0.5 to 99.5 percentiles range and normalized to zero mean and unit deviation. During the training stage, we introduce an on-the-fly data augmentation, with a possibility of $50\%$ in each epoch, by randomly applying the horizontal and vertical flipping, rotation with an angle varying in $[-\pi, \pi]$, perspective distortion with a scale of 0.2, and a Gaussian Blur with a kernel size of $3\time 3$.
\subsection{Evaluation Metrics}\label{sec:metrics}
We evaluate the performance of the DL-based segmentation through four primary metrics in terms of the overlap-based ones, i.e., Dice Similarity Coefficient (DSC) and sensitivity (SEN), spatial-distance-based metric, i.e., Hausdorff distance (HD), and volume-based metric, i.e., volumetric similarity (VS).

The time required for labeling is also measured. Each record of duration begins with the annotation of the first slice of a volume and ends upon completion of the final slice. The duration encompasses not only the labeling but also includes necessary activities such as saving, tool switching, and slice switching.
\subsection{Results}\label{sec:results}
% \subsubsection{Inference in State-Of-The-Art DL-based models}\label{sec:re1}
\subsubsection{Results in local dataset}\label{sec:re1}
\begin{table*}[h]
\setlength{\tabcolsep}{0.7mm}
\caption{Results on local dataset, where the performance comparison is from various 2D/3D DL-based models trained by strong labels ($s$), pseudo labels ($p$), and pseudo labels including external dataset MSD ($p+p^\dag$). The performance is presented regarding the labeling time and four evaluation metrics. The result is illustrated for each metric as the mean and standard deviation of the 3-fold cross-validation. The difference between the results of pseudo-label-based and strong-label-based training is shown in the parentheses. For each metric, red highlights the best value across all models.}
   \begin{center}
  \begin{tabular}{cccccccc}
  \hline
  &Model   &\makecell{Label\\type} &\makecell{Labeling\\time (h)} &DSC (\%)$\uparrow $&SEN (\%)$\uparrow $&HD$\downarrow $ &VS (\%)$\uparrow $\\
  \hline
   \multirow{9}{*}{2D}&\multirow{3}{*}{\makecell{Attention U-Net\\\cite{oktay2018attention}}}&$s$&84.6&86.5\scriptsize$\pm$2.1&89.1\scriptsize$\pm$3.1&8.45\scriptsize$\pm$1.90&92.5\scriptsize$\pm$1.3\\
                     ~&         ~ &$p$       &15.2&88.3{\scriptsize$\pm$2.0} \textbf{($\uparrow $1.8)}&89.6{\scriptsize$\pm$2.2} \textbf{($\uparrow $0.5)}&6.78{\scriptsize$\pm$1.45} \textbf{($\downarrow $1.67)}&94.2{\scriptsize$\pm$1.0} \textbf{($\uparrow $1.7)}\\                                         
                     ~&         ~ &$p+p^\dag$&28.5&89.3{\scriptsize$\pm$1.4} \textbf{($\uparrow $2.8)}&90.1{\scriptsize$\pm$1.9} \textbf{($\uparrow $1.0)}&6.72{\scriptsize$\pm$1.05} \textbf{($\downarrow $1.73)}& 94.8{\scriptsize$\pm$0.4} \textbf{($\uparrow $2.3)}\\

  \cline{2-8}
                     ~&\multirow{3}{*}{\makecell{TransUNet\\\cite{chen2021transunet} }}&$s$&84.6&87.1\scriptsize$\pm$2.3&88.8\scriptsize$\pm$2.0&8.46\scriptsize$\pm$0.62&93.4\scriptsize$\pm$1.2\\
                     ~&         ~ &$p$       &15.2&88.9{\scriptsize$\pm$1.6} \textbf{($\uparrow $1.8)}&89.1{\scriptsize$\pm$1.0} \textbf{($\uparrow $0.3)}&5.90{\scriptsize$\pm$0.68} \textbf{($\downarrow $2.56)}&94.7{\scriptsize$\pm$0.6} \textbf{($\uparrow $1.3)}\\
                     ~&         ~ &$p+p^\dag$&28.5&\textcolor{red}{89.6{\scriptsize$\pm$1.1}} \textbf{($\uparrow $2.5)}&\textcolor{red}{90.6{\scriptsize$\pm$2.1}} \textbf{($\uparrow $1.8)}&\textcolor{red}{5.37{\scriptsize$\pm$0.23}} \textbf{($\downarrow $3.09)}&94.8{\scriptsize$\pm$0.7} \textbf{($\uparrow $1.4)} \\

  \cline{2-8}
                     ~&\multirow{3}{*}{\makecell{SwinUNETR\\\cite{hatamizadeh2021swin} }}&$s$&84.6&86.5\scriptsize$\pm$1.5&87.6\scriptsize$\pm$3.0&12.43\scriptsize$\pm$1.67&93.3\scriptsize$\pm$1.1\\
                     ~&         ~ &$p$       &15.2&87.9{\scriptsize$\pm$2.4} \textbf{($\uparrow $1.4)}&89.2{\scriptsize$\pm$1.3} \textbf{($\uparrow $1.6)}&7.60{\scriptsize$\pm$2.18} \textbf{($\downarrow $4.83)}&93.9{\scriptsize$\pm$1.5} \textbf{($\uparrow $0.6)}\\
                     ~&         ~ &$p+p^\dag$&28.5&88.8{\scriptsize$\pm$2.2} \textbf{($\uparrow $2.3)}&89.5{\scriptsize$\pm$2.6} \textbf{($\uparrow $1.9)}&6.65{\scriptsize$\pm$1.63} \textbf{($\downarrow $5.78)}&94.6{\scriptsize$\pm$1.3} \textbf{($\uparrow $1.3)} \\
  \hline
  \multirow{6}{*}{3D}&\multirow{3}{*}{\makecell{3D U-Net\\\cite{cciccek20163d}}}&$s$&84.6&86.6\scriptsize$\pm$2.0&87.3\scriptsize$\pm$3.8&13.45\scriptsize$\pm$3.81&92.9\scriptsize$\pm$1.7\\
                     ~&         ~ &$p$       &15.2&88.0{\scriptsize$\pm$1.5} \textbf{($\uparrow $1.4)}&88.6{\scriptsize$\pm$1.6} \textbf{($\uparrow $1.3)}&7.36{\scriptsize$\pm$1.38} \textbf{($\downarrow $6.09)}&94.0{\scriptsize$\pm$1.0} \textbf{($\uparrow $1.1)}\\
                     ~&         ~ &$p+p^\dag$&28.5&89.3{\scriptsize$\pm$1.6} \textbf{($\uparrow $2.7)}&90.0{\scriptsize$\pm$1.8} \textbf{($\uparrow $2.7)}&6.27{\scriptsize$\pm$1.03} \textbf{($\downarrow $7.18)}&\textcolor{red}{95.1{\scriptsize$\pm$0.6}} \textbf{($\uparrow $2.2)} \\
  \cline{2-8}
                     ~&\multirow{3}{*}{\makecell{TransBTS\\\cite{wang2021transbts} }}&$s$&84.6&85.1\scriptsize$\pm$0.5&86.6\scriptsize$\pm$1.9&18.85\scriptsize$\pm$6.97&91.8\scriptsize$\pm$0.8\\
                     ~&         ~ &$p$       &15.2&86.7{\scriptsize$\pm$1.9} \textbf{($\uparrow $1.6)}&87.9{\scriptsize$\pm$0.7} \textbf{($\uparrow $1.3)}&10.07{\scriptsize$\pm$2.56} \textbf{($\downarrow $8.78)}&93.2{\scriptsize$\pm$1.0} \textbf{($\uparrow $1.4)}\\
                     ~&         ~ &$p+p^\dag$&28.5&88.5{\scriptsize$\pm$0.8} \textbf{($\uparrow $3.4)}&88.9{\scriptsize$\pm$1.7} \textbf{($\uparrow $2.3)}&7.81{\scriptsize$\pm$0.91} \textbf{($\downarrow $10.96)}&94.4{\scriptsize$\pm$0.5} \textbf{($\uparrow $2.6)} \\

\hline
  \end{tabular}
  \end{center}
  \label{tab:ex1}
\end{table*}
\begin{figure*}[!t]
\centering
\begin{minipage}[c]{0.95\linewidth}
\includegraphics[width=\textwidth]{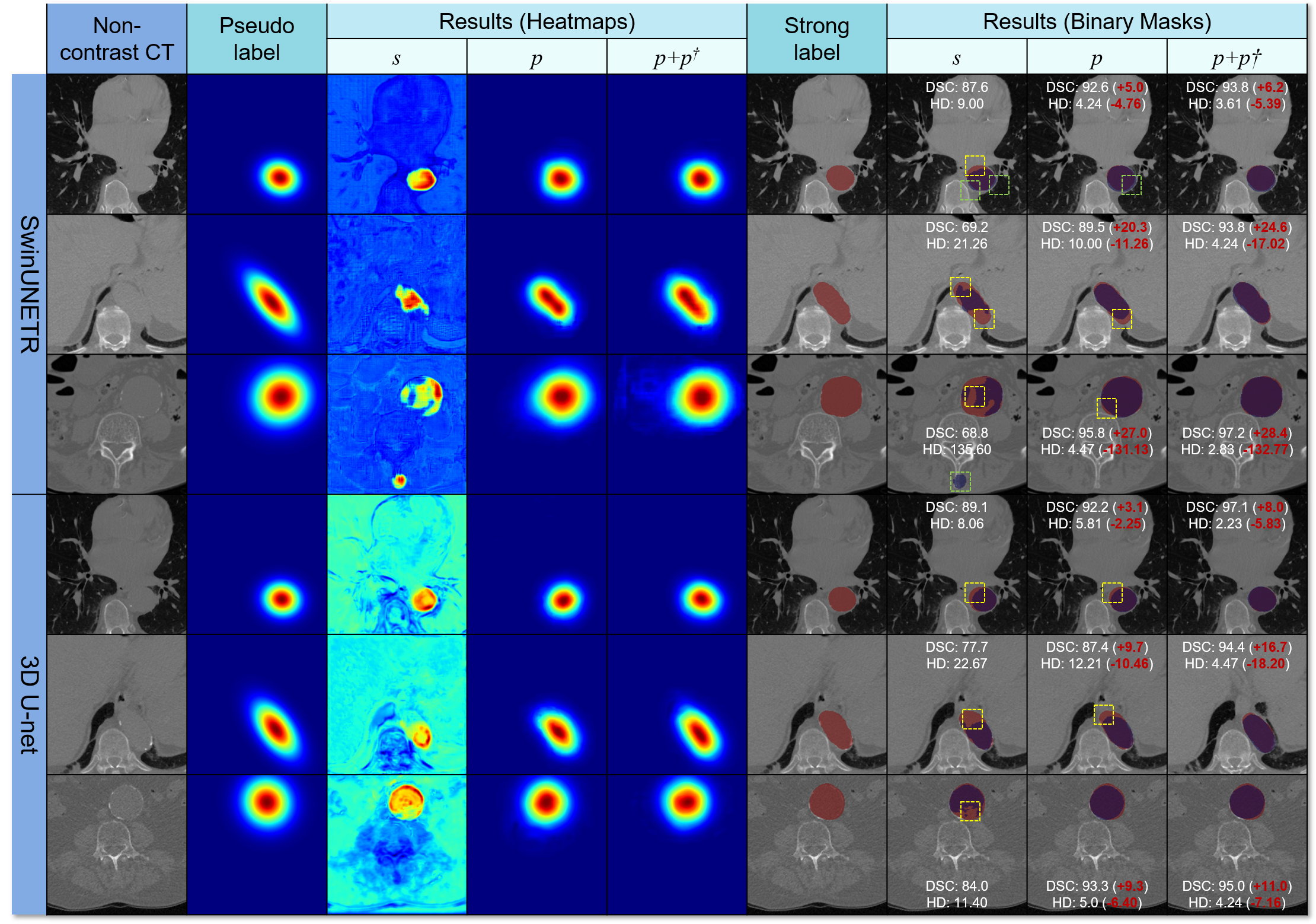}
\end{minipage}
\caption{Visualization of the 2D segmentation results on local data set in terms of heatmap and binary masks from 2D and 3D DL-based models trained by strong labels ($s$), pseudo labels ($p$), and pseudo labels including external dataset MSD ($p+p^\dag$). In binary masks, the red and blue regions (overlaps exhibiting purple) represent the strong labels (ground truths) and predicted results, respectively. For each model, three types of cross sections of the aortas are presented from top to bottom, i.e., the regular-shaped (circular), irregular-shaped (elliptical), and large-sized (aneurysm-contained) ones. The green- and yellow-dotted boxes denote the False Positives (FPs) and False Negatives (FNs), respectively. The Dice coefficient similarity (DSC (\%)) and Hausdorff distance (HD) related to each slice are attached. For each case, the red number in parentheses represents the difference compared to its strong-label-based counterpart ($s$).}\label{fig:s4_2}
\end{figure*}
We initially employed our pseudo-label-based weakly-supervised learning framework in our local dataset through various 2D/3D DL-based models. Table~\ref{tab:ex1} compares the performance of the pseudo-label-based learning and its counterpart of strong-label-based fully-supervised learning. It is observed that pseudo-label-based weakly-supervised learning approaches ($p$) achieve superior performance compared to their strong-label-based counterparts ($s$) for each metric across various 2D/3D models. Performance improvement is accompanied by saving 82.0\% of the labeling time (15.2h vs 84.6h). Introducing the external dataset MSD with pseudo labels ($p+p^\dag$) generates a better performance compared to the counterparts of $p$ with a higher requirement of labeling time of 28.5h, which is still 66.3\% less than the labeling time of strong labels $s$ (84.6h). It is worth noting that, despite the large amount of reduction of labeling time, the pseudo-label-based learning approaches always outperform the strong-label counterparts across various DL-based models, whether with external data or not.

Figure~\ref{fig:s4_2} illustrates the visualization of the randomly sampled 2D results in terms of heatmap and binary masks from a 2D (SwinUNETR \cite{hatamizadeh2021swin}) and a 3D (3D U-Net \cite{cciccek20163d}) DL-based models trained by different types of labels. We observe that each case performs relatively well in the regular-shaped (circular) cross-sectional aorta. However, for the irregular-shaped (elliptical) and large-sized (aneurysm-contained) cross-sectional aortas, the strong-label-trained model yields poor results containing a lot of False Positives (FPs) and False Negatives (FNs), while the pseudo-label-trained models still perform well by predicting the similar distributions explicitly illustrated in the pseudo labels (by observing the pseudo labels and results of the heatmaps). The differences between the visualization of cases of $p/p+p^\dag$ and $s$ qualitatively suggest that the pseudo-label-based weakly-supervised approach achieves superior performance through preserving the topologies of the aortas while alleviating the effects of the FPs and FNs. The improvement of the DSC and HD in each case indicates that the quantitative results are consistent with the qualitative observations.
\begin{figure*}[!t]
\centering
\begin{minipage}[c]{0.95\linewidth}
\includegraphics[width=\textwidth]{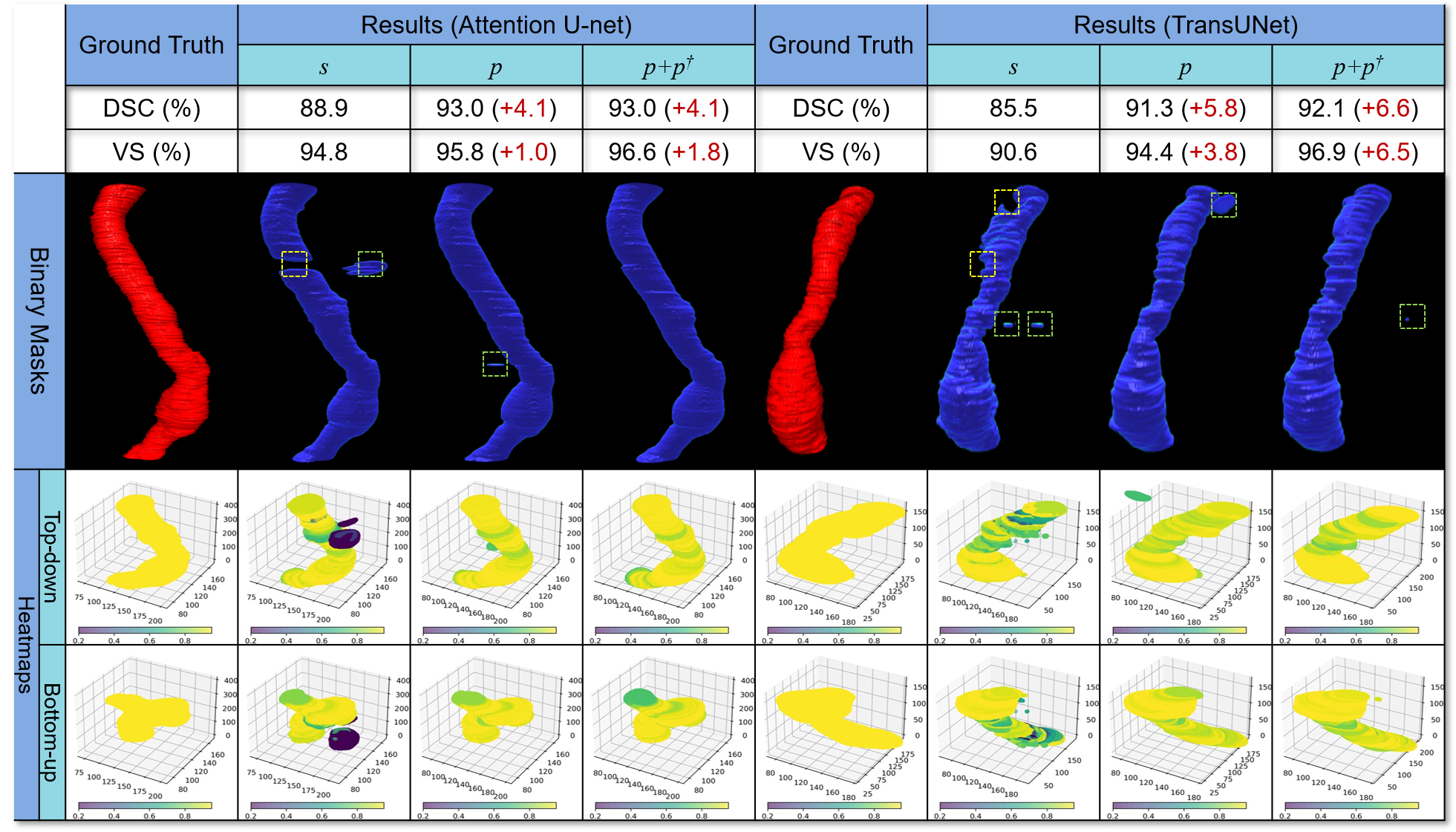}
\end{minipage}
\caption{Visualization of the 3D segmentation results on local data set in terms of binary masks and heatmap from Attention U-net \cite{oktay2018attention} and TransUNet \cite{chen2021transunet} trained by strong labels ($s$), pseudo labels ($p$), and pseudo labels including external dataset MSD ($p+p^\dag$). For each result, the Dice coefficient similarity (DSC) and the volumetric similarity (VS) of the whole volume are presented in the top rows. The green- and yellow-dotted boxes denote the False Positives (FPs) and False Negatives (FNs), respectively. In the results of the heatmaps, the color intensity of each slice correspond to the Dice score of the slice, a lighter color represents a higher Dice score while a darker color means the Dice score of the related slice is lower. The heatmaps are illustrated as top-down and bottom perspectives.}\label{fig:s4_3}
\end{figure*}

Figure~~\ref{fig:s4_3} shows the 3D results in in terms of binary masks and heatmaps. It is observed that the strong-label-trained methods yield obvious FPs and FNs ($s$ of both Attention U-net and TransUNet), which corresponds to the dark regions in the related heatmaps. The pseudo-label-trained methods improves the performance by reducing the majority of FPs and FNs while keep the structure of the abdominal aortas, which is also indicated by the improved DSC and VS.
\subsubsection{Results in TotalSegmentator}\label{sec:re2}
\begin{figure}[!t]
\centering
\begin{minipage}[c]{1.0\linewidth}
\includegraphics[width=\textwidth]{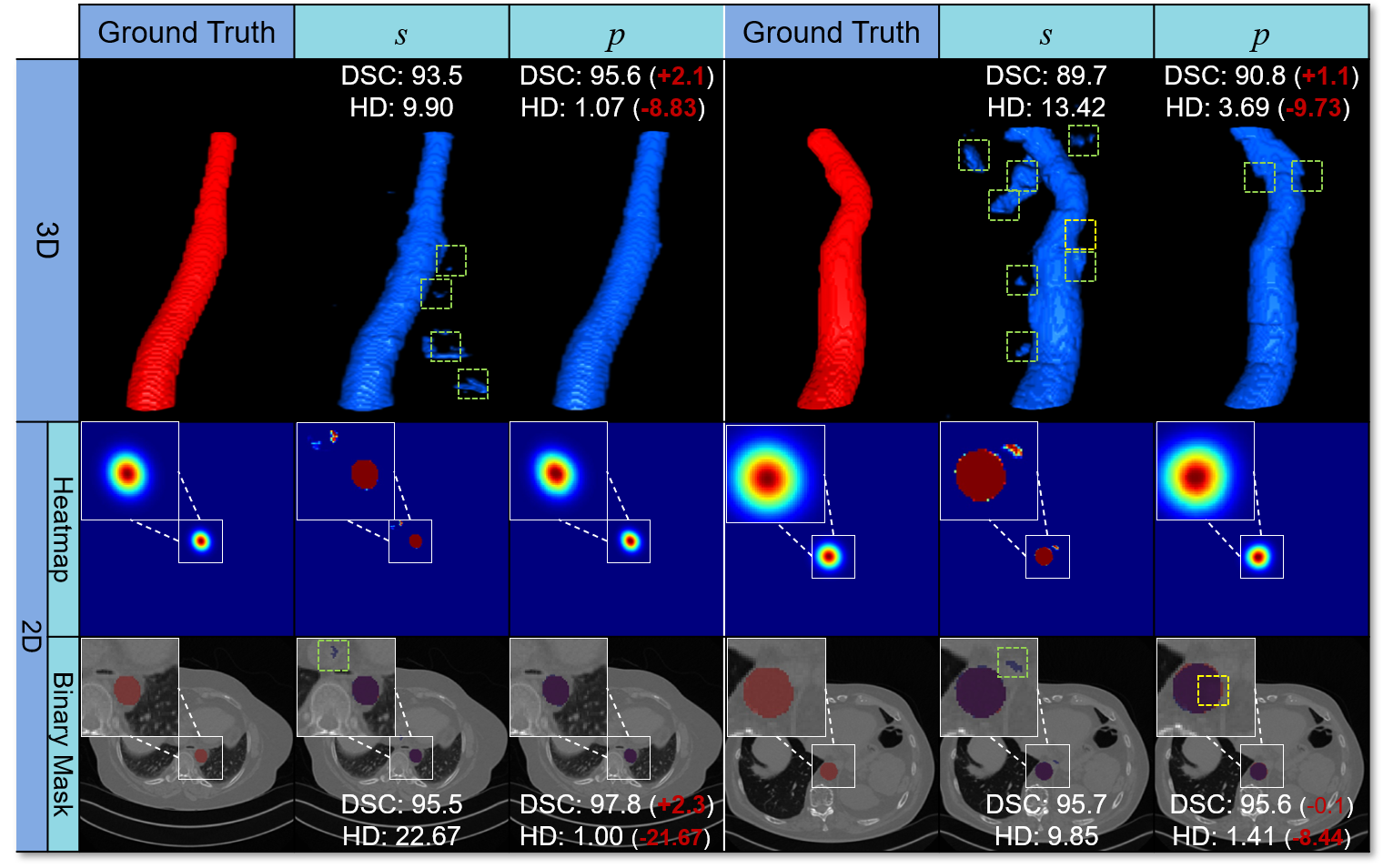}
\end{minipage}
\caption{Visualization of 3D and 2D segmentation results on TotalSegmentator from 3D U-net trained by strong labels ($s$) and pseudo labels ($p$). The visualization of 2D results are displayed in terms of heatmaps and binary masks. The heatmaps are processed by an activation of sigmoid function to highlight the False Positives. In binary masks, the red and blue regions (overlaps exhibiting purple) represent the strong labels (ground truths) and predicted results, respectively. The green- and yellow-dotted boxes denote the False Positives and False Negatives, respectively. The Dice coefficient similarity (DSC (\%)) and Hausdorff distance (HD) related to each volume or slice are attached. For each case, the red number in parentheses represents the difference compared to its strong-label-based counterpart ($s$).}\label{fig:s4_3_ts}
\end{figure}

To evaluate the generalization capability of the proposed method, we conducted analogous experiments on TotalSegmentator. A key distinction is that the pseudo labels were directly derived from the strong labels in this experiment (step (b) and (c) of Figure~\ref{fig:s3_3}), eliminating the need for any additional manual annotation time. Table \ref{tab:totalSeg} shows the performance of 2D/3D models utilizing different labels on the TotalSegmentator dataset. Similar to the results of local dataset, pseudo labels present superior performance over strong labels across various metrics, except for the Dice score of 3D U-net. Notably, pseudo labels yield significantly better results than strong labels in terms of HD (5.41 vs 17.06, 68\% reduced) in 3D U-net. This outcome is comprehensible from Figure \ref{fig:s4_3_ts}. In the 3D visualizations of the results from 3D U-net depicted in Figure \ref{fig:s4_3_ts}, despite their comparable Dice scores, strong labels lead to more FPs. Pseudo labels, by preserving the topology of the aorta, effectively eliminate these FPs, resulting in a more superior HD. The same situation is also observable in their 2D visualizations.

Different from the experiments on local dataset, the experiments on TotalSegmentator specifically focus on the scenarios where strong labels are already provided, exploring the rationality of the conversion of these strong labels into pseudo labels.
\begin{table}[h]
\setlength{\tabcolsep}{0.8mm}
\caption{Results on TotalSegmentator. The performance comparison are from various DL-based models trained by strong labels ($s$) and pseudo labels ($p$). The difference between the results of pseudo-label-based and strong-label-based training is shown in the parentheses.}
   \begin{center}
  \begin{tabular}{ccccccc}
  \hline
  Model                                      &~& DSC (\%)$\uparrow $&SEN (\%)$\uparrow $&HD$\downarrow $ &VS (\%)$\uparrow $\\
  \hline
   \multirow{2}{*}{\makecell{Attention\\U-Net}}&$s$ &85.5&90.2&5.59&89.3\\
   % \cline{2-6}
                              ~ &$p$&88.1 \textbf{($\uparrow $2.6)}&94.9 \textbf{($\uparrow $4.7)}&3.88 \textbf{($\downarrow $1.71)}&90.2 \textbf{($\uparrow $0.9)}\\
\hline
   \multirow{2}{*}{\makecell{Swin-\\UNETR}}&$s$ &83.8&89.3&6.52&87.1\\
   % \cline{2-6}
                              ~ &$p$&85.1 \textbf{($\uparrow $1.3)}&92.4 \textbf{($\uparrow $3.1)}&4.20 \textbf{($\downarrow $2.32)}&88.1 \textbf{($\uparrow $1.0)}\\
                    \hline
   \multirow{2}{*}{\makecell{3D\\U-net}}&$s$ &87.6&91.9&17.06&90.2\\
   % \cline{2-6}
                              ~ &$p$&87.0 ($\downarrow $0.6)&92.5 \textbf{($\uparrow $0.6)}&5.41 \textbf{($\downarrow $11.65)}&90.6 \textbf{($\uparrow $0.4)}\\
\hline
  \end{tabular}
  \end{center}
  \label{tab:totalSeg}
\end{table}

\subsection{Ablation Studies}\label{sec:as}
To further validate the proposed method, we conducted a series of ablation experiments, primarily focusing on the local dataset and the external dataset MSD. These datasets were chosen because they contain more complex structures (e.g., AAAs present in local data), and are more challenging to be processed.
\subsubsection{Performance in relation to the number of training data and pre-train}\label{sec:as}
\begin{figure}[!t]
\centering
\begin{minipage}[c]{1\linewidth}
\includegraphics[width=1.0\textwidth]{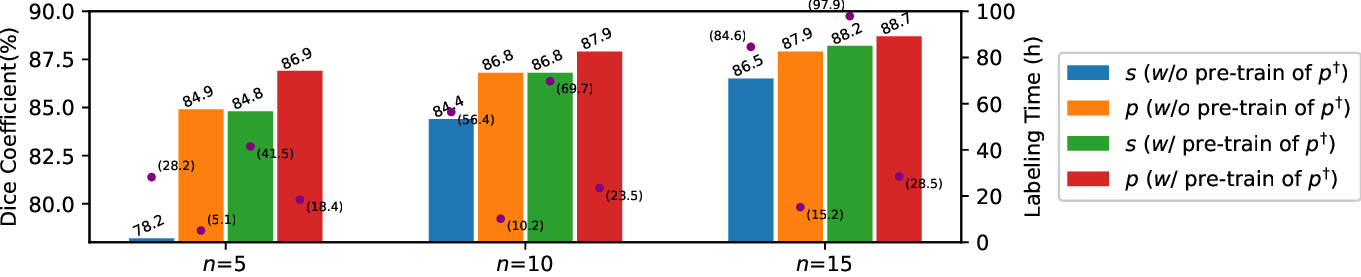}
\end{minipage}
\caption{Performance of SwinUNETR in relation to the number of training data $n$ and to pre-train through external data with pseudo label $p^\dag$. The $y$ coordinates on the left and right sides are Dice score and labeling time, respectively. The bars represent the Dice scores with the numerical values on their top. The purple points mean the labeling time related to each case, with the numerical values in the parentheses.}\label{fig:s4_4}
\end{figure}
To evaluate the performance of strong and pseudo labels in relation to the number of training data, we conduct experiments with train sets containing data volume $n$ ($n = \left\{ {5,10,15} \right\}$) in the DL model SwinUNETR \cite{hatamizadeh2021swin}. Meanwhile, we explore another conventional way of leveraging the external data with pseudo label $p^\dag$, i.e., using it for pre-training the models. 

Figure~\ref{fig:s4_4} shows the performance of different numbers of strong and pseudo labels, with and without the pre-train of external data. In each group of the three, by comparing the orange with blue and the red with the green bars, respectively, we observe that the pseudo-label-based models achieve superior performance than the strong label counterparts, especially in the occasion of extreme lack of training data ($n=5$), where the pseudo labels improve the Dice score by 6.7\% (84.9 vs 78.2) and 2.1\% (86.9 vs 84.8), respectively.

By comparing the green and blue, the red and yellow bars in each group, it is observed that pre-train with external data improves the performance of downstream tasks in both strong- and pseudo-label-based training. For example, in the case of $n=5$, strong-label-based training, the pre-train improves the Dice score by 6.6\% (84.8 vs 78.2). It is worth noting that since the pre-train of external data is based on pseudo labels, it does not cost a high expenditure of additional labeling time (13.3h), less than half the time of labeling strong labels of 5 cases (28.2h).

Comparing the labeling time in each group, for the non-pre-trained cases, the annotation of pseudo labels reduces the labeling time of strong labels by 82\%. For the pre-trained cases, the labeling time is saved by 55.7\%, 66.2\%, and 70.8\%, respectively, which increases with the amount of strong label data.
\subsubsection{Effects of fine-tuning of strong labels}\label{sec:ef}
\begin{figure}
  \begin{minipage}[t]{0.48\linewidth}
    \centering
    \includegraphics[width=\textwidth]{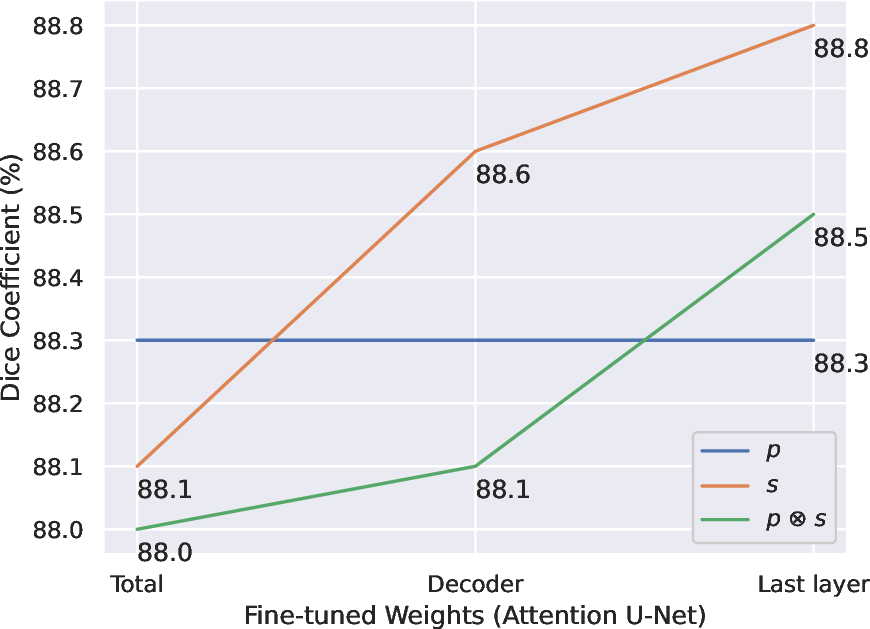}
  \end{minipage}%
   \hspace{0.1in}
  \begin{minipage}[t]{0.48\linewidth}
    \centering
    \includegraphics[width=\textwidth]{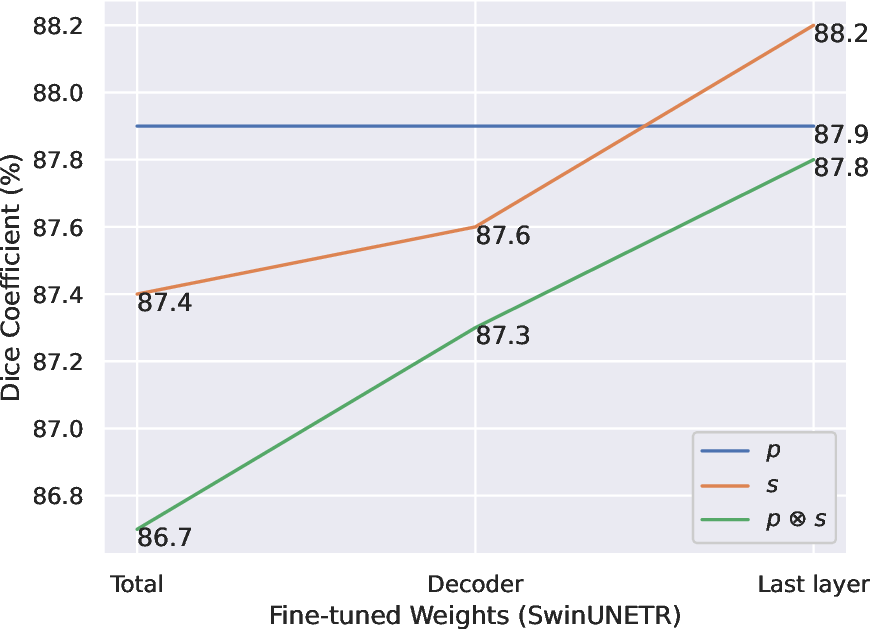}
  \end{minipage}
  \caption{Results of fine-tuning different parts of the pseudo-label ($p$) pre-trained models through the two strong labels ($s$ and $p \otimes s$). The DL models are Attention U-net (left) and SwinUNETR (right). The blue lines represent the results of pre-train with the pseudo labels $p$, and the orange and green lines represent the results of the fine-tuning of the label $s$ and $p \otimes s$, respectively. The $s$ represents the original strong labels while $p \otimes s$ is the element-wise multiplication of $p$ and $s$. The x-axis means fine-tuning various parts of the models, where 'Total' is fine-tuning all the weights of the model while 'Decoder' and 'Last layer' mean fine-tuning the decoder and the last convolutional layer of the models, respectively.
}\label{fig:s4_5_0}
\end{figure}

In this section, we explore a strategy for leveraging both pseudo and strong labels in a unified training task. Intuitively, pseudo-label-based learning merely preserves the topology of an aorta while neglecting its distribution of boundary, which is only presented in a strong label and may not be a perfect conic section. To introduce the knowledge of the pattern of boundaries, we use two types of strong labels to perform the fine-tuning for the pseudo-label pre-trained models, respectively. One is the original strong label $s$ while the other is $p \otimes s$, the element-wise multiplication of $p$ and $s$. The two strong labels are used with the convention 'Dice and BCE' and the proposed 'distribution and reconstruction' loss functions in the fine-tuning stage, respectively. Figure~\ref{fig:s4_5_0} shows the results of fine-tuning different parts of the pseudo-label pre-trained models through the two strong labels. It presents that fine-tuning the last convolutional layer generates the best performance in terms of the Dice score in both DL models, where the $s$-based fine-tuning achieves a better Dice score than $p \otimes s$.

Therefore, we further explore this case in terms of various metrics, recording the results in Table~\ref{tab:ex2}. The pre-training is conducted with two types of labels, i.e., $p$ and $p'$, with $p'$ representing the binary elliptical structures obtained from the efficient manual labeling (Figure~\ref{fig:s3_3} (a)). Across the metrics, $p$ outperforms $p'$, with the most pronounced improvements observed in SEN and HD. Regarding fine-tuning, it shows that the statistical significance (p-value $p<0.05$) of the difference between the results of fine-tuning and pre-training is mainly found in the DSC and SEN. For $p'$-based pre-traning, the $s$-based fine-tuning ($s$ $(p')$) improves the SEN ($1.2\%$ and $2.3\%$, $p<0.05$). However, its overall performance falls short of the results achieved by fine-tuning with $p$-based pretraining ($s$ $(p)$ and $p \otimes s$ $(p)$), especially in terms of DSC and SEN. For $p$-based pre-traning, in Attention U-Net, the fine-tuning of $s$ and $p \otimes s$ achieves a tiny improvement with statistical significance ($0.5\%$ and $0.2\%$, $p<0.05$) in DSC; while in SwinUNETR, the statistical significance is shown in SEN. It is worth noting that based on pre-traning on $p$, $p \otimes s$ achieves a significant improvement of SEN in both DL models (1.7\% and 1.6\%, $p<0.05$), where $s$ reduces the performance in terms of this metric. 

\begin{figure*}[!t]
\centering
\begin{minipage}[c]{0.85\linewidth}
\includegraphics[width=\textwidth]{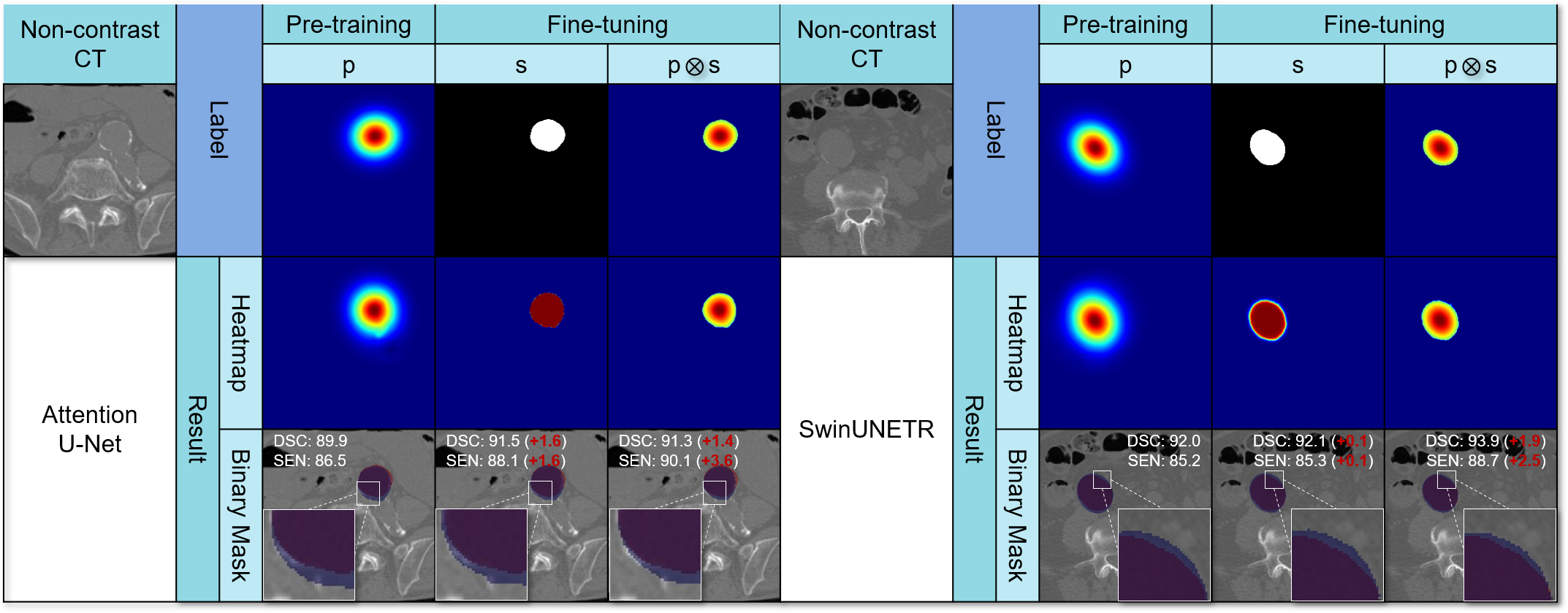}
\end{minipage}
\caption{Visualization of the labels and results of pre-training and fine-tuning stages. '$p$' represents the pseudo label for pre-train, $s$ and $p \otimes s$ are the strong labels for the fine-tuning, where '$s$' means the original strong label and '$p \otimes s$' is the element-wise multiplication of $p$ and $s$. The Dice coefficient similarity (DSC (\%)) and Sensitivity (SEN (\%)) related to each slice are attached. For each case in $s$ and $p \otimes s$, the red number in parentheses represents the difference compared to its pre-trained counterpart ($p$). The white boxes are zoomed in for better observation.}\label{fig:s4_5}
\end{figure*}

To qualitatively perceive this variation, we visualize the results of $p$-based pre-training and $s$/$p \otimes s$-based fine-tuning in Figure~\ref{fig:s4_5}. It is observed that the $p$-based pre-training approaches generate outputs as Gaussian-like distributions without explicit boundaries (before thresholding) while the fine-tuning achieves certain boundaries since the label $s$ and $p \otimes s$ provide these patterns. Comparing the results of binary masks (after thresholding of 0.5), the fine-tuning of $s$ and $p \otimes s$ improve the SEN by reducing the FPs, where $p \otimes s$ eliminates more FPs and achieves higher SENs, which is suggested to have statistical significance according to the Table~\ref{tab:ex2}.

\begin{table*}[h]
\caption{Results of fine-tuning the last convolutional layer. $p$ and $p'$ are used for pre-training, where $p$ means the pseudo labels while $p'$ represents the elliptical structure obtained from the efficient manual labeling (Figure~\ref{fig:s3_3} (a)), used as binary masks. $s$ and $p \otimes s$ are for the fine-tuning, where $s$ represents the original strong labels while $p \otimes s$ is the element-wise multiplication of $p$ and $s$. In ‘Fine-tuning’, \textbf{X (Y)} means the label \textbf{X} is used the for fine-tuning the weights which are pre-trained by \textbf{Y}. 'Sample' illustrates the examples of the visualization of the labels. 'Dice+BCE' represents the conventional loss function while 'KL+MAE' means the combination of Kullback–Leibler (KL) divergence loss and Mean Absolute Error (MAE) loss proposed in Section~\ref{sec:loss}. For results, the increase or decrease in parentheses is relative to its corresponding pre-training. For each metric, bold indicates the best value within the model, while red highlights the best value across all models. The values with '*' indicate the statistically significant difference from its corresponding pre-training, with p-values less than 0.05, implemented by pairwise Wilcoxon Rank Sum Test.}
   \begin{center}
  \begin{tabular}{cccccccc}
  \hline
  Model                                       & Stage &Label&Sample &Loss& DSC (\%)$\uparrow $&SEN (\%)$\uparrow $&HD$\downarrow $ \\
  \hline
   \multirow{5}{*}{\makecell{Attention U-Net}}&\multirow{2}{*}{Pre-train}&
   $p'$  &\raisebox{-.3\height}{\includegraphics[width=0.03\linewidth]{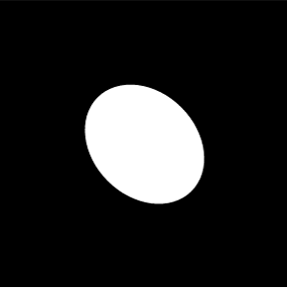}}&Dice+BCE&87.2{\scriptsize$\pm$1.7}&86.8{\scriptsize$\pm$1.1}&8.11{\scriptsize$\pm$0.96}\\
   % 87.5\pm1.8&86.8\pm3.0&8.11\pm1.44&93.1\pm1.3
   ~&~&$p$  &\raisebox{-.3\height}{\includegraphics[width=0.03\linewidth]{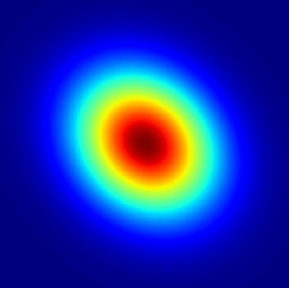}}&KL+MAE&88.3{\scriptsize$\pm$2.0}&89.6{\scriptsize$\pm$2.2}&6.78{\scriptsize$\pm$1.45}\\
   
   \cline{2-8}
   
   ~ &\multirow{3}{*}{\makecell{Fine-\\tuning}}& $s$ $(p')$ &\raisebox{-.3\height}{\includegraphics[width=0.03\linewidth]{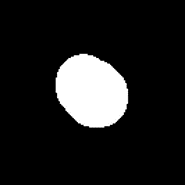}}  &Dice+BCE&87.8{\scriptsize$\pm$1.8}* \textbf{($\uparrow $0.6)}&88.0{\scriptsize$\pm$2.3}* (\textbf{$\uparrow $1.2})&7.79 {\scriptsize$\pm$1.41} (\textbf{$\downarrow $0.32})\\
~ &~ &$s$ $(p)$ &\raisebox{-.3\height}{\includegraphics[width=0.03\linewidth]{s4_7_s.png}}  &Dice+BCE&\textcolor{red}{\textbf{88.8{\scriptsize$\pm$2.0}}*} \textbf{($\uparrow $0.5)}&89.2{\scriptsize$\pm$2.4} ($\downarrow $0.4)&6.90 {\scriptsize$\pm$1.59} ($\uparrow $0.12)\\
   ~ &~ &$p \otimes s$ $(p)$ &\raisebox{-.3\height}{\includegraphics[width=0.03\linewidth]{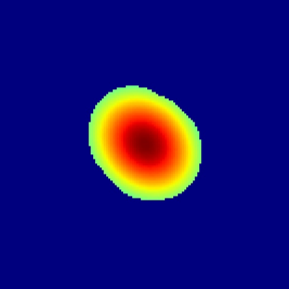}} &KL+MAE&88.5{\scriptsize$\pm$1.6}* \textbf{($\uparrow $0.2)}&\textcolor{red}{\textbf{91.3{\scriptsize$\pm$2.5}}*} \textbf{($\uparrow $1.7)}&\textcolor{red}{\textbf{6.78{\scriptsize$\pm$1.34}}}\\                                   
  
  \hline
  
  \multirow{5}{*}{\makecell{SwinUNETR}}
  &\multirow{2}{*}{Pre-train}&
   $p'$  &\raisebox{-.3\height}{\includegraphics[width=0.03\linewidth]{s4_7_p1.png}}&Dice+BCE&86.6{\scriptsize$\pm$2.6}&86.7{\scriptsize$\pm$1.4}&10.35{\scriptsize$\pm$1.88}\\
  ~&~&$p$ &\raisebox{-.3\height}{\includegraphics[width=0.03\linewidth]{s4_7_p.png}}  &KL+MAE&87.9{\scriptsize$\pm$2.4}&89.2{\scriptsize$\pm$1.3}&\textbf{7.60{\scriptsize$\pm$2.18}}\\
   \cline{2-8}
  ~ &\multirow{3}{*}{\makecell{Fine-\\tuning}}& $s$ $(p')$ &\raisebox{-.3\height}{\includegraphics[width=0.03\linewidth]{s4_7_s.png}}  &Dice+BCE&87.1{\scriptsize$\pm$2.2} \textbf{($\uparrow $0.5)}&88.5{\scriptsize$\pm$3.3}* (\textbf{$\uparrow $1.8})&9.15 {\scriptsize$\pm$2.81} (\textbf{$\downarrow $1.20})\\
   ~&~&$s$ $(p)$&\raisebox{-.3\height}{\includegraphics[width=0.03\linewidth]{s4_7_s.png}}  &Dice+BCE&\textbf{88.2{\scriptsize$\pm$2.2}} \textbf{($\uparrow $0.3)}&88.8{\scriptsize$\pm$2.4}* ($\downarrow $0.4)&7.99{\scriptsize$\pm$2.49} ($\uparrow $0.39)\\
   ~ &~ &$p \otimes s$ $(p)$&\raisebox{-.3\height}{\includegraphics[width=0.03\linewidth]{s4_7_ps.png}}   &KL+MAE&87.8{\scriptsize$\pm$2.3} ($\downarrow $0.1)&\textbf{90.8{\scriptsize$\pm$2.1}}* \textbf{($\uparrow $1.6)}&7.77{\scriptsize$\pm$2.16} ($\uparrow $0.17)\\
\hline
  \end{tabular}
  \end{center}
  \label{tab:ex2}
\end{table*}
\subsubsection{Performance in relation to the loss functions and labels}\label{sec:loss}
We conduct experiments with various loss functions and labels to evaluate their effects. Table~\ref{tab:loss1} shows the results of the 2D and 3D DL models trained with local pseudo label $p$ through various combinations of loss functions. It is observed that the combination of the distribution loss and reconstruction loss generates superior performance for both 2D and 3D models, where the reconstruction loss is Mean Absolute Error (MAE) loss, and the distribution loss is Kullback–Leibler (KL) divergence loss in 2D while Wasserstein loss in 3D model. The KL loss is inappropriate for the 3D model because it only considers the distribution of the aorta in a cross-section, where the contextual information between the slices is ignored. The inapplicability is also present in using Wasserstein loss independently for both models, which means that considering only the Expectation $\mathbb{E}$ of the difference between the output and pseudo label $p$ does not work in this case. However, the combination with MAE loss improves the performance of MAE-only trained models, which indicates that the distribution loss can overcome the limitation that reconstruction loss only considers the voxel values and ignores their distribution.

Table~\ref{tab:loss2} shows the performance of the 2D and 3D models with various labels and their corresponding loss functions. Overall, we considered four types of labels, including the strong labels $s$, the pseudo labels $p$, the element-wise multiplications $p \otimes s$, and the binary ellipse-like structures $p'$. The labels $s$ and $p'$ are binary masks trained with conventional Dice and BCE loss, while the $p$ and $p \otimes s$ are trained with the combination of distribution and reconstruction loss. It is observed that $p$ generates superior performance compared to other labels, where KL and Wasserstein loss are optimal for the 2D and 3D models, respectively. Compared to the results of $s$-based training, $p$ shows improvement with statistically significant differences across all models and metrics. The $p'$, however, as an intermediate form of $s$ and $p$, showed a significant decrease (1.0\%) in SEN with TransUnet and did not exhibit significant improvement in DSC and HD with 3D U-Net. The $p \otimes s$ achieves inferior results compared to $p$, considering the studies in the Section~\ref{sec:ef}, it is suggested that the $p \otimes s$ is more applicable to be leveraged in the fine-tuning stage if the strong labels are offered. It is worth noticing that the labeling time of $p$ and $p'$ are all 15.2h, reducing 82\% of the labeling time of $s$ (84.6h), while the labeling time of $p \otimes s$ is the sum of the labeling time of $p$ and $s$ (99.8h) since both the labels are involved.

\begin{table*}[h]
\setlength{\tabcolsep}{0.7mm}
\caption{Results of the 2D and 3D models trained with pseudo label $p$. The loss functions are the Mean Absolute Error (MAE) loss, Kullback-Leibler (KL) divergence loss, Wasserstein loss, and their combinations. The bold and underlined values indicate the column's optimal and sub-optimal values, respectively. '/' means the loss function is inapplicable to the pseudo label $p$.}
   \begin{center}
  \begin{tabular}{ccccccc}
  \hline
    ~   & \multicolumn{3}{c}{TransUNet (2D)} & \multicolumn{3}{c}{3D U-net (3D)}\\
  
   Loss & DSC (\%)$\uparrow $&SEN (\%)$\uparrow $&HD$\downarrow $ & DSC (\%)$\uparrow $&SEN (\%)$\uparrow $&HD$\downarrow $\\
  \hline
   MAE& 88.3{\scriptsize$\pm$1.3}&88.6{\scriptsize$\pm$0.9}&6.58{\scriptsize$\pm$0.73}&\underline{87.7{\scriptsize$\pm$1.6}}&\underline{88.1{\scriptsize$\pm$2.2}}&\underline{7.60{\scriptsize$\pm$1.47}}\\
   Wass&  &/& &  &/& \\
   KL& 88.0{\scriptsize$\pm$1.6}&88.1{\scriptsize$\pm$1.8}&\underline{6.33{\scriptsize$\pm$1.09}}& &/&\\
   Wass+MAE& \underline{88.5{\scriptsize$\pm$1.8}}&\underline{89.0{\scriptsize$\pm$1.5}}&6.36{\scriptsize$\pm$0.97}&\textbf{88.0{\scriptsize$\pm$1.5}}&\textbf{88.6{\scriptsize$\pm$1.6}}&\textbf{7.36{\scriptsize$\pm$1.38}}\\
   KL+MAE& \textbf{88.9{\scriptsize$\pm$1.6}}&\textbf{89.1{\scriptsize$\pm$1.0}}&\textbf{5.90{\scriptsize$\pm$0.68}}&&/&\\                
\hline
  \end{tabular}
  \end{center}
  \label{tab:loss1}
\end{table*}

\begin{table*}[h]
\setlength{\tabcolsep}{0.5mm}
\caption{Results of the 2D and 3D models trained with various labels and loss functions. The four types of labels are strong labels $s$, the binary ellipse-like structures $p'$, the pseudo labels $p$, and the element-wise multiplications $p \otimes s$. 'Sample' illustrates the examples of the visualization of the labels. The binary mask labels ($s$ and $p'$) are trained with the combinations of the Dice and BCE loss, while the heatmap labels ($p$ and $p \otimes s$) are trained with the combination of distribution loss (KL or Wass) and reconstruction loss (MAE). The bold and underlined values indicate the column's optimal and sub-optimal values, respectively. '/' means the loss function is inapplicable to the label.The values with '*' indicate the statistically significant difference from the results of $s$, with p-values less than 0.05, implemented by pairwise Wilcoxon Rank Sum Test.}
   \begin{center}
  \begin{tabular}{cccccccccc}
  \hline
    \multirow{2}{*}{Label}&\multirow{2}{*}{Sample}&\multirow{2}{*}{\makecell{Labeling\\time (h)}}&\multirow{2}{*}{Loss} & \multicolumn{3}{c}{TransUNet (2D)} & \multicolumn{3}{c}{3D U-net (3D)}\\
  
  ~&~&~&~& DSC (\%)$\uparrow $&SEN (\%)$\uparrow $&HD$\downarrow $ & DSC (\%)$\uparrow $&SEN (\%)$\uparrow $&HD$\downarrow $\\
  \hline
   $s$ &\raisebox{-.5\height}{\includegraphics[width=0.03\linewidth]{s4_7_s.png}}&84.6&Dice+BCE& 87.1{\scriptsize$\pm$2.3}&88.8{\scriptsize$\pm$2.0}&8.46{\scriptsize$\pm$0.62}&86.6{\scriptsize$\pm$2.0}&87.3{\scriptsize$\pm$3.8}&13.45{\scriptsize$\pm$3.81}\\
   
   $p'$&\raisebox{-.5\height}{\includegraphics[width=0.03\linewidth]{s4_7_p1.png}}&15.2&Dice+BCE& 88.1{\scriptsize$\pm$1.4}*&87.8{\scriptsize$\pm$3.3}* &7.49{\scriptsize$\pm$2.03}*&87.0{\scriptsize$\pm$1.8}&87.6{\scriptsize$\pm$2.1}*&11.02{\scriptsize$\pm$2.18}\\
   
   \hline
   $p$&\multirow{2}{*}{\raisebox{-.5\height}{\includegraphics[width=0.03\linewidth]{s4_7_p.png}}} &15.2&Wass+MAE& \underline{88.5{\scriptsize$\pm$1.8}}*&\underline{89.0{\scriptsize$\pm$1.5}}*&\underline{6.36{\scriptsize$\pm$0.97}}*&\textbf{88.0{\scriptsize$\pm$1.5}}*&\textbf{88.6{\scriptsize$\pm$1.6}}*&\textbf{7.36{\scriptsize$\pm$1.38}}*\\
   $p$&~ &15.2&KL+MAE  & \textbf{88.9{\scriptsize$\pm$1.6}}*&\textbf{89.1{\scriptsize$\pm$1.0}}*&\textbf{5.90{\scriptsize$\pm$0.68}}*&&/&\\
   \hline
   $p \otimes s$&\multirow{2}{*}{\raisebox{-.5\height}{\includegraphics[width=0.03\linewidth]{s4_7_ps.png}}}&99.8&Wass+MAE& 87.7{\scriptsize$\pm$1.2}*&88.8{\scriptsize$\pm$1.3}&6.89{\scriptsize$\pm$1.27}*&\underline{87.2{\scriptsize$\pm$1.3}}*&\underline{87.6{\scriptsize$\pm$1.8}}*&\underline{9.15{\scriptsize$\pm$2.13}}*\\ 
   $p \otimes s$&~&99.8&KL+MAE& 87.9{\scriptsize$\pm$1.5}*&88.9{\scriptsize$\pm$1.2}&6.58{\scriptsize$\pm$0.78}*&&/&\\                
\hline
  \end{tabular}
  \end{center}
  \label{tab:loss2}
\end{table*}
\subsubsection{Generalization in the other anatomical structure}\label{sec:prostate}
To assess the proposed method's generalizability to other ellipse-like anatomical structures, we apply it to prostate segmentation in CTs. We randomly sampled 30 volumes containing prostates from the TotalSegmentator dataset. Similar to our aortic CTs, these prostate CTs are without contrast agents, exhibiting ambiguous boundaries. Additionally, the prostates present ellipse-like shapes in the majority of cross-sections. Data preprocessing was conducted as described in section \ref{subsec:totalseg}, and 3-fold cross-validation followed the local data allocation in Figure~\ref{fig:s4_1}. With the strong label $s$ provided, we automatically generated Gaussian pseudo-labels $p$ and their binarized structures $p'$ as outlined in section \ref{subsec:totalseg}. We leverage TransUNet for training according to its superior performance in aortic segmentation.

\begin{figure}[!t]
\centering
\begin{minipage}[c]{0.95\linewidth}
\includegraphics[width=\textwidth]{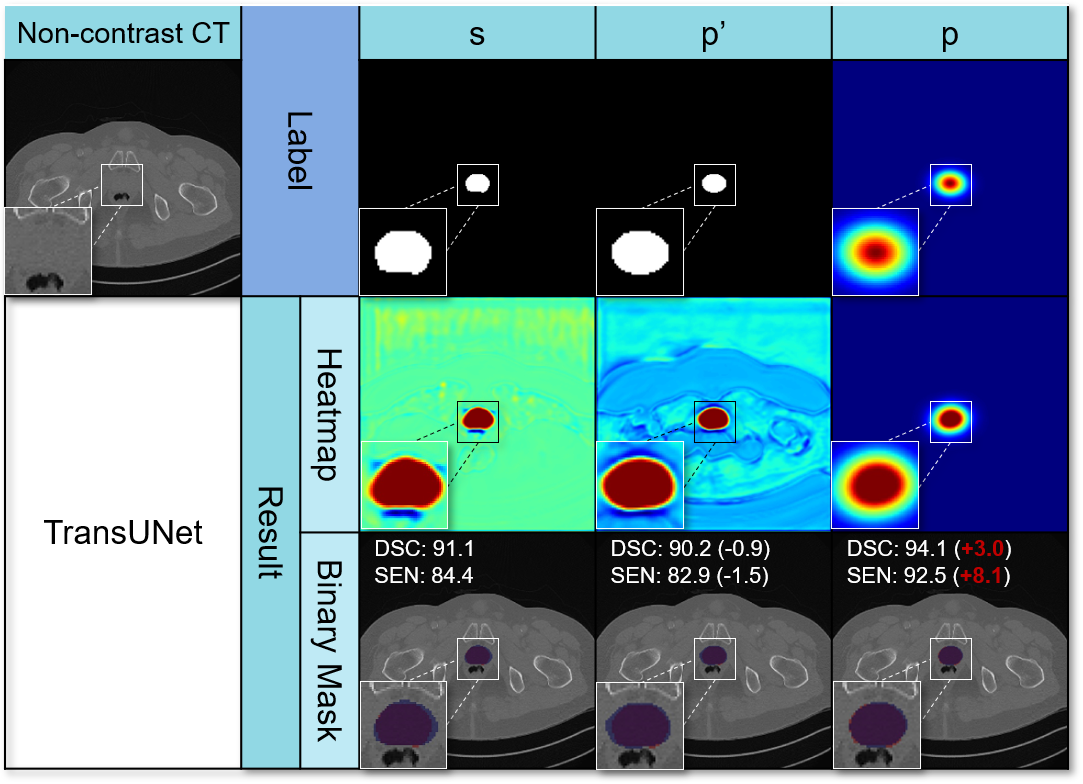}
\end{minipage}
\caption{Visualization of the labels and results in terms of heatmap and binary masks of prostate obtained from TransUNet. It is trained by strong labels ($s$), binary elliptical structures ($p'$), and pseudo labels ($p$). In the results of binary masks, the red and blue regions (overlaps exhibiting purple) represent the strong labels (ground truths) and predicted results, respectively. The Dice coefficient similarity (DSC (\%)) and Sensitivity (SEN(\%)) related to each slice are attached. For each case in $p'$ and $p$, the number in parentheses represents the difference compared to $s$, red indicates the improvement. The boxes within the slices are zoomed in for better observation.}\label{fig:s5}
\end{figure}

Table~\ref{tab:prostate} presents the results of training with different labels. We observed that $p$ obtains performance improvements over $s$ across all metrics, with statistically significant differences, particularly in SEN and HD. However, the binary elliptical structure $p'$ showed significant decreases in DSC and HD, with no significant change in VS. Figure~\ref{fig:s5} visualizes the different labels and their corresponding results. Results based on binary labels ($s$ and $p'$) exhibited obvious false positives. Although $p'$, as an elliptical structure, partially preserved the topological form, its errors in boundary regions led to declines in both metrics. In contrast, $p$ effectively maintained the topology while controlling boundary extensions, resulting in better performance.
\begin{table}[h]
\setlength{\tabcolsep}{0.7mm}
\caption{Results of prostate on Totalsegmentator. The performance comparison is from TransUNet trained by strong labels ($s$), binary elliptical structures ($p'$), and pseudo labels ($p$). The result is illustrated for each metric as the mean of the 3-fold cross-validation. The difference between the results of $p$/$p'$ and $s$ is shown in the parentheses. The values with '*' indicate the statistically significant difference from the results of $s$, with p-values less than 0.05, implemented by pairwise Wilcoxon Rank Sum Test.}
   \begin{center}
  \begin{tabular}{ccccc}
  \hline
  Label &DSC (\%)$\uparrow $&SEN (\%)$\uparrow $&HD$\downarrow $ &VS (\%)$\uparrow $\\
  \hline
   $s$&87.0&85.1&8.79&94.5\\
   \cline{2-5}
   $p'$&86.7* ($\downarrow $0.3)&86.2* \textbf{($\uparrow $1.1)}&10.20* ($\uparrow $1.41)&94.3 ($\downarrow$0.2)\\                                         
   $p$&\textbf{87.9}* \textbf{($\uparrow $0.9)}&\textbf{88.6}* \textbf{($\uparrow $3.5)}&\textbf{4.13}* \textbf{($\downarrow $4.66)}&\textbf{95.1}* \textbf{($\uparrow $0.6)}\\
\hline
  \end{tabular}
  \end{center}
  \label{tab:prostate}
\end{table}
\section{Discussion} \label{sec:dis}
We proposed a weakly-supervised learning approach for the segmentation of ellipse-like VS in non-contrast CTs. It focuses on the abdominal aorta based on the Gaussian-like pseudo labels. The generation of pseudo labels consists of (1) efficient labeling based on the proposed annotation standards, (2) ellipse fitting, and (3) Gaussian heatmap generation. The pseudo labels are integrated into the DL models' training through a novel combination of voxel reconstruction and distribution losses. The experimental results exhibited their effectiveness.

In comparison to the three pseudo-label types presented in Section \ref{sec:related_pseudo_labels} (CAMs, sparse, and noisy labels), our pseudo-labels incorporate characteristics from all three. The Gaussian heatmaps are a particular type of CAMs; the manually annotated ellipse-like structures represent a sparsification of strong labels, saving labeling time; and retaining topological features without explicit boundaries can be regarded as a form of noisy label. All these elements contribute to the effectiveness of our pseudo labels.

Table~\ref{tab:ex1} demonstrates the superiority of our approach from both the perspectives of reducing annotation time and enhancing model performance. The reduction in annotation time is attributed to a series of proposed annotation standards, which resulted in an 82\% reduction in local data annotation time while mitigating the excessive reliance on the supervision of surgeons. Despite the substantial reduction in annotation time, the model's performance significantly improved across various metrics, surpassing the strong-label-based fully-supervised learning models. We suggest it could also outperform the conventional weakly supervised learning methods that must navigate a trade-off between annotation time and accuracy. We attribute the enhancement in model performance to the Gaussian-based pseudo labels, which reflect the general characteristics of the VS in CT slices. This approach preserves the topological nature of the aorta while avoiding the ambiguous impact of boundaries in non-contrast CT scans. The decrease in HD substantiates our hypothesis. Meanwhile, owing to the introduced annotation standards, the intra-/inter-observer consistency of pseudo labels ensures the stability of model training. These factors collectively contribute to the improvement in model performance. The substantial reduction in annotation time facilitates the introduction of external data. Consequently, our approach can be applied to label and generate pseudo labels with minimal annotation costs for a label-agnostic external dataset, further enhancing performance ($p+p^\dag$ in Table~\ref{tab:ex1}). 

Table~\ref{tab:totalSeg} illustrates the effectiveness of the proposed method on a public dataset TotalSegmentator, where the pseudo labels are directly derived from strong labels without incurring any manual annotation time. Considering both Table~\ref{tab:ex1} and Table~\ref{tab:totalSeg}, the former addresses label-agnostic datasets, showing the competitiveness of our model through efficient labeling and pseudo label generation. The latter deals with datasets containing strong labels, revealing the rationality of converting them into pseudo labels to enhance model performance without incurring additional annotation time. 

The qualitative results in Figures~\ref{fig:s4_2}-~\ref{fig:s4_3_ts} align with the quantitative findings in Tables~\ref{tab:ex1}-~\ref{tab:totalSeg}. As shown in Figures~\ref{fig:s4_2}, the performance of the strong label \( s \) is weak in segmenting aortic cross-sections that are irregular or of large size, which are mainly related to the aneurysmal parts. We observed that the aneurysmal regions exhibit greater variability compared to non-aneurysmal areas, and this variability manifests as increased discrepancies between the training and test sets. We hypothesize the variability increases difficulty in segmenting the aneurysmal regions. In contrast, the pseudo label \( p \) yields superior performance, as shown in Figures~\ref{fig:s4_2}, with most false positives (FP) and false negatives (FN) being eliminated. We infer that the primary advantage of the pseudo label \( p \) over the strong label \( s \) lies in the supervision signal of Gaussian-based consistency. Although the supervision signals may differ in shape and size, models can capture the consistency, enabling them to better handle varying aortic cross-sections. As a result, the pseudo label outperforms the strong label, particularly in irregular and large aortic cross-sections.

The ablation studies made further explorations. Section~\ref{sec:as} shows a strategy for leveraging external data, which can be utilized for pre-training to provide prior knowledge for downstream segmentation tasks. Additionally, as depicted in Figure~\ref{fig:s4_4}, it reveals that pseudo labels outperform strong labels in scenarios with minimal training data ($n$=5). We posit that this superior performance of pseudo labels is attributed to their better preservation of the general characteristics of the VS in slices. We argue that this general nature facilitates the model to mitigate overfitting. This is analogous to traditional machine learning classification, where models are often trained on the most prominent features of samples to achieve a smooth decision boundary between different classes, rather than overfitting to noisy features. Similarly, our pseudo labels retain the general characteristics of the samples, such as smooth boundaries and ellipse-like topology, while removing potential noisy features, such as complex boundaries. This overfitting mitigation becomes increasingly evident as the amount of training data decreases. As the dataset size increases, the overfitting associated with strong labels tends to diminish due to the greater diversity of samples. However, the substantial time required for labeling with strong labels becomes a significant challenge. In contrast, while the advantages of pseudo label derived from prior ellipse-like knowledge may decrease, the efficiency gains in labeling costs become more pronounced. Therefore, when dealing with larger datasets, balancing performance and labeling time emerges as an open issue.

Section \ref{sec:ef} explores a combined use of strong labels and pseudo labels, i.e., fine-tuning a model trained on pseudo labels with the addition of strong labels. We observed a significant improvement in Sensitivity when using $p \otimes s$ as strong labels for fine-tuning with KL+MAE loss (Table~\ref{tab:ex2}), compared to the pre-training stage. We attribute this improvement to the consistency of the loss functions during fine-tuning and pre-training. Figure~\ref{fig:s4_5} further illustrates this achievement by showing a reduction in false positives near the boundaries, highlighting the superior sensitivity enhancement of $p \otimes s$ over $s$.

Section~\ref{sec:loss} investigates the impact of different loss functions and labels on performance. Table~\ref{tab:loss1} indicates that our proposed combinations of reconstruction and distribution losses are more beneficial for training with pseudo labels. We observed that the proposed combination of KL+MAE for 2D and Wasserstein+MAE yielded stability during training. The instability occurs exclusively in the ablation study. The unsuitability of KL in the 3D scenario is attributed to the increased complexity of the distribution to be fitted, which led to gradient instability. The ineffectiveness of using Wasserstein individually is due to the slow gradient updates, resulting in gradient vanishing. Table~\ref{tab:loss2} shows that Gaussian-based pseudo labels ($p$) exhibit the most robust performance among the four types of labels. By comparing labels $s$, $p$, and $p'$, we assume that labels with a more "general" morphology ($p$ and $p'$) achieve superior performance. Furthermore, $p$ outperforms $p'$. We attribute the superiority to the smoothness, continuity, and informational richness of the Gaussian heatmap. The smoothness and continuity facilitate the model’s understanding of target boundaries and regions through smooth and continuous transitions. The informational richness extends beyond binary information to include the probability densities of the target. This detailed information aids the model in better localization and recognition of the target during prediction.

This paper focuses on the abdominal aorta with and without aneurysms in non-contrast CTs. Given that the axial cross-sections of the abdominal aorta (or aneurysm) can be approximated by ellipses, the annotation process is directly conducted along the axial CT slices. However, not all VS can be adequately fitted with a single ellipse in their axial cross-sections. Considering different types of VS is still required for the generalization of our pseudo-labeling approach. Examples include the aortic arch (non-convex form), ascending/descending aorta (dual objects), and iliac arteries (dual objects with small sizes). For these structures, an alternative annotating approach along the trajectory of the VS, within the cross-sections perpendicular to the VS centerline, becomes a feasible strategy. This should ensure the ellipse-like topology of the VS in the cross-sections.

A potential interest of our method is its ability to enhance the competitiveness of non-contrast CT relative to CTA, particularly considering the medical benefits for the patients associated with non-contrast imaging and the performance of our method on such data. When compared to state-of-the-art methods \cite{vagenas2023deep} on the CTA dataset AVT \cite{radl2022avt}, our approach achieved results of a comparable magnitude on non-contrast CTs, even with fewer training data. Therefore, we suggest that this method could increase the usability of non-contrast CTs in clinical applications. With the enhancement of such methods, non-contrast CT is expected to gain higher priority in routine diagnostic procedures

\begin{figure}[!t]
\centering
\begin{minipage}[c]{0.95\linewidth}
\includegraphics[width=\textwidth]{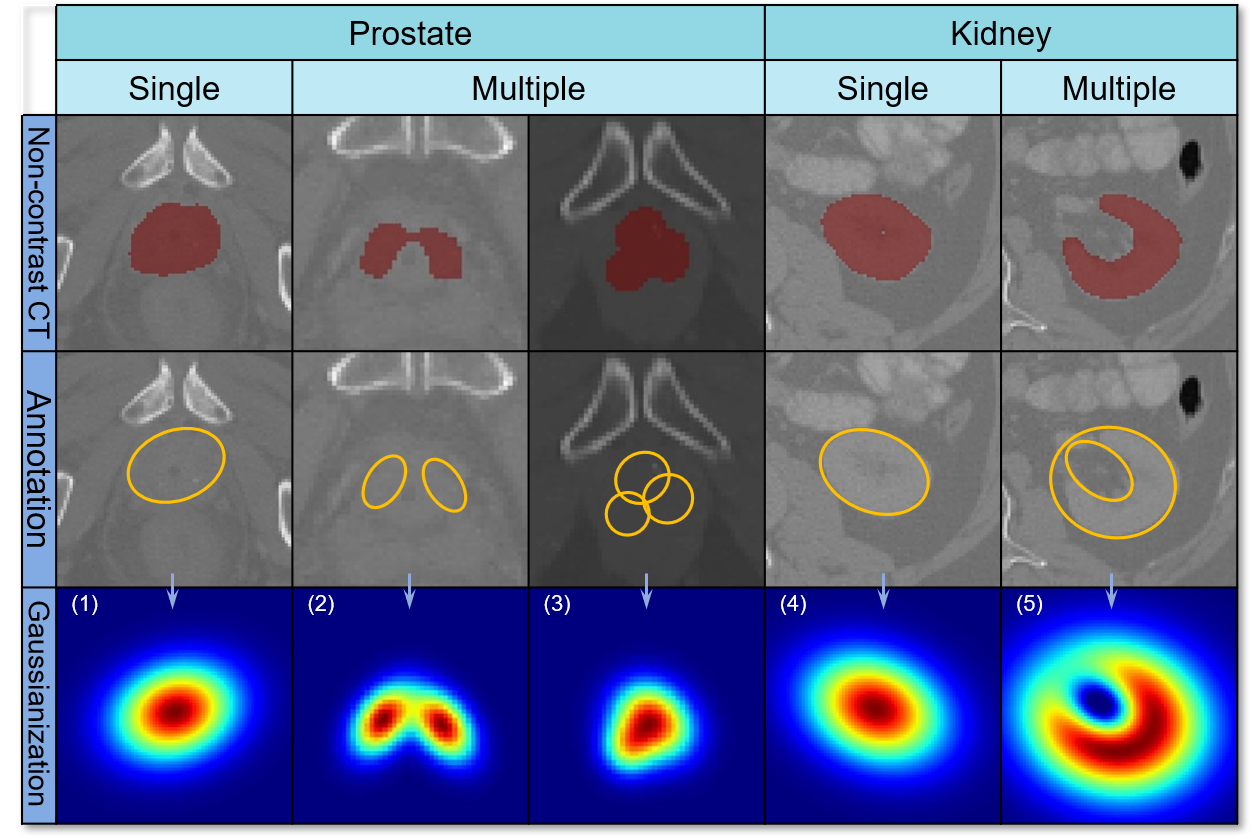}
\end{minipage}
\caption{Generalization of Gaussian pseudo labels to other anatomical structures. The first row of the visualization shows non-contrast CTs and the ground truth of the anatomical structures (in red). The second row shows the efficient annotation method using elliptical structures, modeling the target with either single or multiple ellipses. The third row illustrates the generated pseudo labels. Single elliptical structures ((1) and (4)) are Gaussianized as shown in Figure~\ref{fig:s3_3} (b) and (c), while multiple elliptical structures are Gaussianized individually and then combined through summation ((2) and (3)) or subtraction (5), followed by normalization. This process creates overlapping spatial Gaussian distributions that can simulate the target anatomical structure besides aorta.}\label{fig:diss}
\end{figure}

For anatomical structures beyond the aorta, the target structure can be approximated using overlapping spatial Gaussian distributions. Figure~\ref{fig:diss} illustrates examples of the prostate and kidney. Some of them can be represented by a single Gaussian distribution ((1) and (4)), while more complex structures can be modeled by summation ((2) and (3)) or subtraction (5) of multiple spatial Gaussian distributions. Combining with an appropriate loss function, these pseudo labels have potential to be generalized to target anatomical structures to reduce labeling costs.
\section{Conclusion} \label{sec:con}
The DL-based segmentation of VS in non-contrast CTs faces challenges of extensive data annotation and substantial intra-/inter-observer variability due to ambiguous boundaries in non-contrast CTs. The former consumes excessive annotation time and relies heavily on surgical supervision, while the latter diminishes model performance and stability. This paper addresses these challenges in the context of the abdominal aorta, a typical vascular structure. We propose a weakly-supervised learning framework that leverages the elliptical approximation of the abdominal aorta's topological form in CT slices. Gaussian heatmaps, generated from the best-fitted ellipses of the aortas, are utilized as pseudo labels. The proposed annotation standards significantly reduce annotation time, while the Gaussian heatmaps preserve the intrinsic characteristics of the abdominal aorta and mitigate the negative impact of ambiguous boundaries, enhancing model stability and performance. Experiments conducted on both label-agnostic datasets (local data and MSD) and labeled datasets (TotalSegmentator) demonstrate the effectiveness of our approach. Pseudo labels outperform strong labels in terms of both annotation time and model performance. Future research will extend this methodology to different types of vascular structures.
\section*{Acknowledgments}
This study was partially supported by the French National Research Agency (ANR) in the framework of the Investissement d’Avenir Program through Labex CAMI (ANR-11- LABX-0004). This work was supported in part by the National Key Research and Development Program of China under Grant 2022YFE0116700. The first author is grateful for the support of China Scholarship Council (CSC Grant No. 201906090389)
% \section*{References}

% Please ensure that every reference cited in the text is also present in
% the reference list (and vice versa).

% \section*{\itshape Reference style}

% Text: All citations in the text should refer to:
% \begin{enumerate}
% \item Single author: the author's name (without initials, unless there
% is ambiguity) and the year of publication;
% \item Two authors: both authors' names and the year of publication;
% \item Three or more authors: first author's name followed by `et al.'
% and the year of publication.
% \end{enumerate}
% Citations may be made directly (or parenthetically). Groups of
% references should be listed first alphabetically, then chronologically.

% %%Harvard
\bibliographystyle{model2-names.bst}\biboptions{authoryear}
\bibliography{refs}

\begin{thebibliography}{60}
\expandafter\ifx\csname natexlab\endcsname\relax\def\natexlab#1{#1}\fi
\providecommand{\url}[1]{\texttt{#1}}
\providecommand{\href}[2]{#2}
\providecommand{\path}[1]{#1}
\providecommand{\DOIprefix}{doi:}
\providecommand{\ArXivprefix}{arXiv:}
\providecommand{\URLprefix}{URL: }
\providecommand{\Pubmedprefix}{pmid:}
\providecommand{\doi}[1]{\href{http://dx.doi.org/#1}{\path{#1}}}
\providecommand{\Pubmed}[1]{\href{pmid:#1}{\path{#1}}}
\providecommand{\bibinfo}[2]{#2}
\ifx\xfnm\relax \def\xfnm[#1]{\unskip,\space#1}\fi
%Type = Inproceedings
\bibitem[{Ahn et~al.(2019)Ahn, Cho and Kwak}]{ahn2019weakly}
\bibinfo{author}{Ahn, J.}, \bibinfo{author}{Cho, S.}, \bibinfo{author}{Kwak,
  S.}, \bibinfo{year}{2019}.
\newblock \bibinfo{title}{Weakly supervised learning of instance segmentation
  with inter-pixel relations}, in: \bibinfo{booktitle}{Proceedings of the
  IEEE/CVF conference on computer vision and pattern recognition}, pp.
  \bibinfo{pages}{2209--2218}.
%Type = Article
\bibitem[{Antonelli et~al.(2022)Antonelli, Reinke, Bakas, Farahani,
  Kopp-Schneider, Landman, Litjens, Menze, Ronneberger, Summers
  et~al.}]{antonelli2022medical}
\bibinfo{author}{Antonelli, M.}, \bibinfo{author}{Reinke, A.},
  \bibinfo{author}{Bakas, S.}, \bibinfo{author}{Farahani, K.},
  \bibinfo{author}{Kopp-Schneider, A.}, \bibinfo{author}{Landman, B.A.},
  \bibinfo{author}{Litjens, G.}, \bibinfo{author}{Menze, B.},
  \bibinfo{author}{Ronneberger, O.}, \bibinfo{author}{Summers, R.M.}, et~al.,
  \bibinfo{year}{2022}.
\newblock \bibinfo{title}{The medical segmentation decathlon}.
\newblock \bibinfo{journal}{Nature communications} \bibinfo{volume}{13},
  \bibinfo{pages}{4128}.
%Type = Article
\bibitem[{Baldeon-Calisto and Lai-Yuen(2020)}]{baldeon2020adaresu}
\bibinfo{author}{Baldeon-Calisto, M.}, \bibinfo{author}{Lai-Yuen, S.K.},
  \bibinfo{year}{2020}.
\newblock \bibinfo{title}{Adaresu-net: Multiobjective adaptive convolutional
  neural network for medical image segmentation}.
\newblock \bibinfo{journal}{Neurocomputing} \bibinfo{volume}{392},
  \bibinfo{pages}{325--340}.
%Type = Article
\bibitem[{Chandrashekar et~al.(2022)Chandrashekar, Handa, Shivakumar, Lapolla,
  Grau and Lee}]{chandrashekar2020deep}
\bibinfo{author}{Chandrashekar, A.}, \bibinfo{author}{Handa, A.},
  \bibinfo{author}{Shivakumar, N.}, \bibinfo{author}{Lapolla, P.},
  \bibinfo{author}{Grau, V.}, \bibinfo{author}{Lee, R.}, \bibinfo{year}{2022}.
\newblock \bibinfo{title}{A deep learning approach to automate high-resolution
  blood vessel reconstruction on computerized tomography images with or without
  the use of contrast agent}.
\newblock \bibinfo{journal}{Annals of Surgery} \bibinfo{volume}{276},
  \bibinfo{pages}{1017--1027}.
%Type = Article
\bibitem[{Chen et~al.(2021)Chen, Lu, Yu, Luo, Adeli, Wang, Lu, Yuille and
  Zhou}]{chen2021transunet}
\bibinfo{author}{Chen, J.}, \bibinfo{author}{Lu, Y.}, \bibinfo{author}{Yu, Q.},
  \bibinfo{author}{Luo, X.}, \bibinfo{author}{Adeli, E.},
  \bibinfo{author}{Wang, Y.}, \bibinfo{author}{Lu, L.},
  \bibinfo{author}{Yuille, A.L.}, \bibinfo{author}{Zhou, Y.},
  \bibinfo{year}{2021}.
\newblock \bibinfo{title}{Transunet: Transformers make strong encoders for
  medical image segmentation}.
\newblock \bibinfo{journal}{arXiv preprint arXiv:2102.04306} .
%Type = Inproceedings
\bibitem[{{\c{C}}i{\c{c}}ek et~al.(2016){\c{C}}i{\c{c}}ek, Abdulkadir,
  Lienkamp, Brox and Ronneberger}]{cciccek20163d}
\bibinfo{author}{{\c{C}}i{\c{c}}ek, {\"O}.}, \bibinfo{author}{Abdulkadir, A.},
  \bibinfo{author}{Lienkamp, S.S.}, \bibinfo{author}{Brox, T.},
  \bibinfo{author}{Ronneberger, O.}, \bibinfo{year}{2016}.
\newblock \bibinfo{title}{3d u-net: learning dense volumetric segmentation from
  sparse annotation}, in: \bibinfo{booktitle}{Medical Image Computing and
  Computer-Assisted Intervention--MICCAI 2016: 19th International Conference,
  Athens, Greece, October 17-21, 2016, Proceedings, Part II 19},
  \bibinfo{organization}{Springer}. pp. \bibinfo{pages}{424--432}.
%Type = Article
\bibitem[{Davenport et~al.(2013)Davenport, Khalatbari, Dillman, Cohan, Caoili
  and Ellis}]{davenport2013contrast}
\bibinfo{author}{Davenport, M.S.}, \bibinfo{author}{Khalatbari, S.},
  \bibinfo{author}{Dillman, J.R.}, \bibinfo{author}{Cohan, R.H.},
  \bibinfo{author}{Caoili, E.M.}, \bibinfo{author}{Ellis, J.H.},
  \bibinfo{year}{2013}.
\newblock \bibinfo{title}{Contrast material--induced nephrotoxicity and
  intravenous low-osmolality iodinated contrast material}.
\newblock \bibinfo{journal}{Radiology} \bibinfo{volume}{267},
  \bibinfo{pages}{94--105}.
%Type = Article
\bibitem[{Foley and Karcaaltincaba(2003)}]{foley2003computed}
\bibinfo{author}{Foley, W.D.}, \bibinfo{author}{Karcaaltincaba, M.},
  \bibinfo{year}{2003}.
\newblock \bibinfo{title}{Computed tomography angiography: principles and
  clinical applications}.
\newblock \bibinfo{journal}{Journal of computer assisted tomography}
  \bibinfo{volume}{27}, \bibinfo{pages}{S23--S30}.
%Type = Article
\bibitem[{Fu et~al.(2023)Fu, Xu, Chang, Yang, Ling, Cai, Chen, Yuan, Cai, Zhang
  et~al.}]{fu2023robust}
\bibinfo{author}{Fu, S.}, \bibinfo{author}{Xu, J.}, \bibinfo{author}{Chang,
  S.}, \bibinfo{author}{Yang, L.}, \bibinfo{author}{Ling, S.},
  \bibinfo{author}{Cai, J.}, \bibinfo{author}{Chen, J.}, \bibinfo{author}{Yuan,
  J.}, \bibinfo{author}{Cai, Y.}, \bibinfo{author}{Zhang, B.}, et~al.,
  \bibinfo{year}{2023}.
\newblock \bibinfo{title}{Robust vascular segmentation for raw complex images
  of laser speckle contrast based on weakly supervised learning}.
\newblock \bibinfo{journal}{IEEE Transactions on Medical Imaging}
  \bibinfo{volume}{43}, \bibinfo{pages}{39--50}.
%Type = Article
\bibitem[{Gu et~al.(2018)Gu, Shen, Yang and Yang}]{gu2018reliable}
\bibinfo{author}{Gu, Y.}, \bibinfo{author}{Shen, M.}, \bibinfo{author}{Yang,
  J.}, \bibinfo{author}{Yang, G.Z.}, \bibinfo{year}{2018}.
\newblock \bibinfo{title}{Reliable label-efficient learning for biomedical
  image recognition}.
\newblock \bibinfo{journal}{IEEE Transactions on Biomedical Engineering}
  \bibinfo{volume}{66}, \bibinfo{pages}{2423--2432}.
%Type = Article
\bibitem[{Guo et~al.(2024)Guo, Tan, Feng and Zhou}]{guo20243d}
\bibinfo{author}{Guo, Z.}, \bibinfo{author}{Tan, Z.}, \bibinfo{author}{Feng,
  J.}, \bibinfo{author}{Zhou, J.}, \bibinfo{year}{2024}.
\newblock \bibinfo{title}{3d vascular segmentation supervised by 2d annotation
  of maximum intensity projection}.
\newblock \bibinfo{journal}{IEEE Transactions on Medical Imaging} .
%Type = Inproceedings
\bibitem[{Hal{\i}r and Flusser(1998)}]{halir1998numerically}
\bibinfo{author}{Hal{\i}r, R.}, \bibinfo{author}{Flusser, J.},
  \bibinfo{year}{1998}.
\newblock \bibinfo{title}{Numerically stable direct least squares fitting of
  ellipses}, in: \bibinfo{booktitle}{Proc. 6th International Conference in
  Central Europe on Computer Graphics and Visualization. WSCG},
  \bibinfo{organization}{Citeseer}. pp. \bibinfo{pages}{125--132}.
%Type = Book
\bibitem[{Haralick and Shapiro(1992)}]{haralick1992computer}
\bibinfo{author}{Haralick, R.M.}, \bibinfo{author}{Shapiro, L.G.},
  \bibinfo{year}{1992}.
\newblock \bibinfo{title}{Computer and robot vision}.
  volume~\bibinfo{volume}{1}.
\newblock \bibinfo{publisher}{Addison-wesley Reading, MA}.
%Type = Inproceedings
\bibitem[{Hatamizadeh et~al.(2021)Hatamizadeh, Nath, Tang, Yang, Roth and
  Xu}]{hatamizadeh2021swin}
\bibinfo{author}{Hatamizadeh, A.}, \bibinfo{author}{Nath, V.},
  \bibinfo{author}{Tang, Y.}, \bibinfo{author}{Yang, D.},
  \bibinfo{author}{Roth, H.R.}, \bibinfo{author}{Xu, D.}, \bibinfo{year}{2021}.
\newblock \bibinfo{title}{Swin unetr: Swin transformers for semantic
  segmentation of brain tumors in mri images}, in:
  \bibinfo{booktitle}{International MICCAI Brainlesion Workshop},
  \bibinfo{organization}{Springer}. pp. \bibinfo{pages}{272--284}.
%Type = Article
\bibitem[{Hinson et~al.(2017)Hinson, Ehmann, Fine, Fishman, Toerper, Rothman
  and Klein}]{hinson2017risk}
\bibinfo{author}{Hinson, J.S.}, \bibinfo{author}{Ehmann, M.R.},
  \bibinfo{author}{Fine, D.M.}, \bibinfo{author}{Fishman, E.K.},
  \bibinfo{author}{Toerper, M.F.}, \bibinfo{author}{Rothman, R.E.},
  \bibinfo{author}{Klein, E.Y.}, \bibinfo{year}{2017}.
\newblock \bibinfo{title}{Risk of acute kidney injury after intravenous
  contrast media administration}.
\newblock \bibinfo{journal}{Annals of emergency medicine} \bibinfo{volume}{69},
  \bibinfo{pages}{577--586}.
%Type = Article
\bibitem[{Isensee et~al.(2021)Isensee, Jaeger, Kohl, Petersen and
  Maier-Hein}]{isensee2021nnu}
\bibinfo{author}{Isensee, F.}, \bibinfo{author}{Jaeger, P.F.},
  \bibinfo{author}{Kohl, S.A.}, \bibinfo{author}{Petersen, J.},
  \bibinfo{author}{Maier-Hein, K.H.}, \bibinfo{year}{2021}.
\newblock \bibinfo{title}{nnu-net: a self-configuring method for deep
  learning-based biomedical image segmentation}.
\newblock \bibinfo{journal}{Nature methods} \bibinfo{volume}{18},
  \bibinfo{pages}{203--211}.
%Type = Article
\bibitem[{Kaladji et~al.(2015)Kaladji, Dumenil, Mah{\'e}, Castro, Cardon, Lucas
  and Haigron}]{kaladji2015safety}
\bibinfo{author}{Kaladji, A.}, \bibinfo{author}{Dumenil, A.},
  \bibinfo{author}{Mah{\'e}, G.}, \bibinfo{author}{Castro, M.},
  \bibinfo{author}{Cardon, A.}, \bibinfo{author}{Lucas, A.},
  \bibinfo{author}{Haigron, P.}, \bibinfo{year}{2015}.
\newblock \bibinfo{title}{Safety and accuracy of endovascular aneurysm repair
  without pre-operative and intra-operative contrast agent}.
\newblock \bibinfo{journal}{European Journal of Vascular and Endovascular
  Surgery} \bibinfo{volume}{49}, \bibinfo{pages}{255--261}.
%Type = Inproceedings
\bibitem[{Khoreva et~al.(2017)Khoreva, Benenson, Hosang, Hein and
  Schiele}]{khoreva2017simple}
\bibinfo{author}{Khoreva, A.}, \bibinfo{author}{Benenson, R.},
  \bibinfo{author}{Hosang, J.}, \bibinfo{author}{Hein, M.},
  \bibinfo{author}{Schiele, B.}, \bibinfo{year}{2017}.
\newblock \bibinfo{title}{Simple does it: Weakly supervised instance and
  semantic segmentation}, in: \bibinfo{booktitle}{Proceedings of the IEEE
  conference on computer vision and pattern recognition}, pp.
  \bibinfo{pages}{876--885}.
%Type = Article
\bibitem[{Kingma and Ba(2014)}]{kingma2014adam}
\bibinfo{author}{Kingma, D.P.}, \bibinfo{author}{Ba, J.}, \bibinfo{year}{2014}.
\newblock \bibinfo{title}{Adam: A method for stochastic optimization}.
\newblock \bibinfo{journal}{arXiv preprint arXiv:1412.6980} .
%Type = Article
\bibitem[{Kullback and Leibler(1951)}]{kullback1951information}
\bibinfo{author}{Kullback, S.}, \bibinfo{author}{Leibler, R.A.},
  \bibinfo{year}{1951}.
\newblock \bibinfo{title}{On information and sufficiency}.
\newblock \bibinfo{journal}{The annals of mathematical statistics}
  \bibinfo{volume}{22}, \bibinfo{pages}{79--86}.
%Type = Article
\bibitem[{LeCun et~al.(2015)LeCun, Bengio and Hinton}]{lecun2015deep}
\bibinfo{author}{LeCun, Y.}, \bibinfo{author}{Bengio, Y.},
  \bibinfo{author}{Hinton, G.}, \bibinfo{year}{2015}.
\newblock \bibinfo{title}{Deep learning}.
\newblock \bibinfo{journal}{nature} \bibinfo{volume}{521},
  \bibinfo{pages}{436--444}.
%Type = Inproceedings
\bibitem[{Lin et~al.(2016)Lin, Dai, Jia, He and Sun}]{lin2016scribblesup}
\bibinfo{author}{Lin, D.}, \bibinfo{author}{Dai, J.}, \bibinfo{author}{Jia,
  J.}, \bibinfo{author}{He, K.}, \bibinfo{author}{Sun, J.},
  \bibinfo{year}{2016}.
\newblock \bibinfo{title}{Scribblesup: Scribble-supervised convolutional
  networks for semantic segmentation}, in: \bibinfo{booktitle}{Proceedings of
  the IEEE conference on computer vision and pattern recognition}, pp.
  \bibinfo{pages}{3159--3167}.
%Type = Article
\bibitem[{Litjens et~al.(2017)Litjens, Kooi, Bejnordi, Setio, Ciompi,
  Ghafoorian, Van Der~Laak, Van~Ginneken and S{\'a}nchez}]{litjens2017survey}
\bibinfo{author}{Litjens, G.}, \bibinfo{author}{Kooi, T.},
  \bibinfo{author}{Bejnordi, B.E.}, \bibinfo{author}{Setio, A.A.A.},
  \bibinfo{author}{Ciompi, F.}, \bibinfo{author}{Ghafoorian, M.},
  \bibinfo{author}{Van Der~Laak, J.A.}, \bibinfo{author}{Van~Ginneken, B.},
  \bibinfo{author}{S{\'a}nchez, C.I.}, \bibinfo{year}{2017}.
\newblock \bibinfo{title}{A survey on deep learning in medical image analysis}.
\newblock \bibinfo{journal}{Medical image analysis} \bibinfo{volume}{42},
  \bibinfo{pages}{60--88}.
%Type = Inproceedings
\bibitem[{Lu et~al.(2019)Lu, Brooks, Hahn, Chen, Buch, Kotecha, Andriole,
  Ghoshhajra, Pinto, Vozila et~al.}]{lu2019deepaaa}
\bibinfo{author}{Lu, J.T.}, \bibinfo{author}{Brooks, R.},
  \bibinfo{author}{Hahn, S.}, \bibinfo{author}{Chen, J.},
  \bibinfo{author}{Buch, V.}, \bibinfo{author}{Kotecha, G.},
  \bibinfo{author}{Andriole, K.P.}, \bibinfo{author}{Ghoshhajra, B.},
  \bibinfo{author}{Pinto, J.}, \bibinfo{author}{Vozila, P.}, et~al.,
  \bibinfo{year}{2019}.
\newblock \bibinfo{title}{Deepaaa: clinically applicable and generalizable
  detection of abdominal aortic aneurysm using deep learning}, in:
  \bibinfo{booktitle}{Medical Image Computing and Computer Assisted
  Intervention--MICCAI 2019: 22nd International Conference, Shenzhen, China,
  October 13--17, 2019, Proceedings, Part II 22},
  \bibinfo{organization}{Springer}. pp. \bibinfo{pages}{723--731}.
%Type = Article
\bibitem[{Ma et~al.(2023)Ma, Lucas, Hammami, Shu, Kaladji and
  Haigron}]{ma2023deep}
\bibinfo{author}{Ma, Q.}, \bibinfo{author}{Lucas, A.},
  \bibinfo{author}{Hammami, H.}, \bibinfo{author}{Shu, H.},
  \bibinfo{author}{Kaladji, A.}, \bibinfo{author}{Haigron, P.},
  \bibinfo{year}{2023}.
\newblock \bibinfo{title}{Deep-learning approach to automate the segmentation
  of aorta in non-contrast cts}.
\newblock \bibinfo{journal}{Journal of Medical Imaging} \bibinfo{volume}{10},
  \bibinfo{pages}{024001--024001}.
%Type = Inproceedings
\bibitem[{Matuszewski and Sintorn(2018)}]{matuszewski2018minimal}
\bibinfo{author}{Matuszewski, D.J.}, \bibinfo{author}{Sintorn, I.M.},
  \bibinfo{year}{2018}.
\newblock \bibinfo{title}{Minimal annotation training for segmentation of
  microscopy images}, in: \bibinfo{booktitle}{2018 IEEE 15th International
  Symposium on Biomedical Imaging (ISBI 2018)}, \bibinfo{organization}{IEEE}.
  pp. \bibinfo{pages}{387--390}.
%Type = Article
\bibitem[{McDonald et~al.(2013)McDonald, McDonald, Bida, Carter, Fleming,
  Misra, Williamson and Kallmes}]{mcdonald2013intravenous}
\bibinfo{author}{McDonald, R.J.}, \bibinfo{author}{McDonald, J.S.},
  \bibinfo{author}{Bida, J.P.}, \bibinfo{author}{Carter, R.E.},
  \bibinfo{author}{Fleming, C.J.}, \bibinfo{author}{Misra, S.},
  \bibinfo{author}{Williamson, E.E.}, \bibinfo{author}{Kallmes, D.F.},
  \bibinfo{year}{2013}.
\newblock \bibinfo{title}{Intravenous contrast material--induced nephropathy:
  causal or coincident phenomenon?}
\newblock \bibinfo{journal}{Radiology} \bibinfo{volume}{267},
  \bibinfo{pages}{106--118}.
%Type = Inproceedings
\bibitem[{Milletari et~al.(2016)Milletari, Navab and Ahmadi}]{milletari2016v}
\bibinfo{author}{Milletari, F.}, \bibinfo{author}{Navab, N.},
  \bibinfo{author}{Ahmadi, S.A.}, \bibinfo{year}{2016}.
\newblock \bibinfo{title}{V-net: Fully convolutional neural networks for
  volumetric medical image segmentation}, in: \bibinfo{booktitle}{2016 fourth
  international conference on 3D vision (3DV)}, \bibinfo{organization}{Ieee}.
  pp. \bibinfo{pages}{565--571}.
%Type = Inproceedings
\bibitem[{Min et~al.(2019)Min, Chen, Zha, Wu and Zhang}]{min2019two}
\bibinfo{author}{Min, S.}, \bibinfo{author}{Chen, X.}, \bibinfo{author}{Zha,
  Z.J.}, \bibinfo{author}{Wu, F.}, \bibinfo{author}{Zhang, Y.},
  \bibinfo{year}{2019}.
\newblock \bibinfo{title}{A two-stream mutual attention network for
  semi-supervised biomedical segmentation with noisy labels}, in:
  \bibinfo{booktitle}{Proceedings of the AAAI Conference on Artificial
  Intelligence}, pp. \bibinfo{pages}{4578--4585}.
%Type = Article
\bibitem[{Minaee et~al.(2021)Minaee, Boykov, Porikli, Plaza, Kehtarnavaz and
  Terzopoulos}]{minaee2021image}
\bibinfo{author}{Minaee, S.}, \bibinfo{author}{Boykov, Y.},
  \bibinfo{author}{Porikli, F.}, \bibinfo{author}{Plaza, A.},
  \bibinfo{author}{Kehtarnavaz, N.}, \bibinfo{author}{Terzopoulos, D.},
  \bibinfo{year}{2021}.
\newblock \bibinfo{title}{Image segmentation using deep learning: A survey}.
\newblock \bibinfo{journal}{IEEE transactions on pattern analysis and machine
  intelligence} \bibinfo{volume}{44}, \bibinfo{pages}{3523--3542}.
%Type = Inproceedings
\bibitem[{Mirikharaji et~al.(2019)Mirikharaji, Yan and
  Hamarneh}]{mirikharaji2019learning}
\bibinfo{author}{Mirikharaji, Z.}, \bibinfo{author}{Yan, Y.},
  \bibinfo{author}{Hamarneh, G.}, \bibinfo{year}{2019}.
\newblock \bibinfo{title}{Learning to segment skin lesions from noisy
  annotations}, in: \bibinfo{booktitle}{Domain Adaptation and Representation
  Transfer and Medical Image Learning with Less Labels and Imperfect Data:
  First MICCAI Workshop, DART 2019, and First International Workshop, MIL3ID
  2019, Shenzhen, Held in Conjunction with MICCAI 2019, Shenzhen, China,
  October 13 and 17, 2019, Proceedings 1}, \bibinfo{organization}{Springer}.
  pp. \bibinfo{pages}{207--215}.
%Type = Inproceedings
\bibitem[{Ning et~al.(2023)Ning, Liang, Chen, Zhang and Liao}]{ning2023doppler}
\bibinfo{author}{Ning, G.}, \bibinfo{author}{Liang, H.}, \bibinfo{author}{Chen,
  F.}, \bibinfo{author}{Zhang, X.}, \bibinfo{author}{Liao, H.},
  \bibinfo{year}{2023}.
\newblock \bibinfo{title}{Doppler image-based weakly-supervised vascular
  ultrasound segmentation with transformer}, in: \bibinfo{booktitle}{2023 IEEE
  20th International Symposium on Biomedical Imaging (ISBI)},
  \bibinfo{organization}{IEEE}. pp. \bibinfo{pages}{1--5}.
%Type = Article
\bibitem[{Oktay et~al.(2018)Oktay, Schlemper, Folgoc, Lee, Heinrich, Misawa,
  Mori, McDonagh, Hammerla, Kainz et~al.}]{oktay2018attention}
\bibinfo{author}{Oktay, O.}, \bibinfo{author}{Schlemper, J.},
  \bibinfo{author}{Folgoc, L.L.}, \bibinfo{author}{Lee, M.},
  \bibinfo{author}{Heinrich, M.}, \bibinfo{author}{Misawa, K.},
  \bibinfo{author}{Mori, K.}, \bibinfo{author}{McDonagh, S.},
  \bibinfo{author}{Hammerla, N.Y.}, \bibinfo{author}{Kainz, B.}, et~al.,
  \bibinfo{year}{2018}.
\newblock \bibinfo{title}{Attention u-net: Learning where to look for the
  pancreas}.
\newblock \bibinfo{journal}{arXiv preprint arXiv:1804.03999} .
%Type = Article
\bibitem[{Paszke et~al.(2019)Paszke, Gross, Massa, Lerer, Bradbury, Chanan,
  Killeen, Lin, Gimelshein, Antiga et~al.}]{paszke2019pytorch}
\bibinfo{author}{Paszke, A.}, \bibinfo{author}{Gross, S.},
  \bibinfo{author}{Massa, F.}, \bibinfo{author}{Lerer, A.},
  \bibinfo{author}{Bradbury, J.}, \bibinfo{author}{Chanan, G.},
  \bibinfo{author}{Killeen, T.}, \bibinfo{author}{Lin, Z.},
  \bibinfo{author}{Gimelshein, N.}, \bibinfo{author}{Antiga, L.}, et~al.,
  \bibinfo{year}{2019}.
\newblock \bibinfo{title}{Pytorch: An imperative style, high-performance deep
  learning library}.
\newblock \bibinfo{journal}{Advances in neural information processing systems}
  \bibinfo{volume}{32}.
%Type = Article
\bibitem[{Power et~al.(2016)Power, Moloney, Twomey, James, O’Connor and
  Maher}]{power2016computed}
\bibinfo{author}{Power, S.P.}, \bibinfo{author}{Moloney, F.},
  \bibinfo{author}{Twomey, M.}, \bibinfo{author}{James, K.},
  \bibinfo{author}{O’Connor, O.J.}, \bibinfo{author}{Maher, M.M.},
  \bibinfo{year}{2016}.
\newblock \bibinfo{title}{Computed tomography and patient risk: Facts,
  perceptions and uncertainties}.
\newblock \bibinfo{journal}{World journal of radiology} \bibinfo{volume}{8},
  \bibinfo{pages}{902}.
%Type = Article
\bibitem[{Radl et~al.(2022)Radl, Jin, Pepe, Li, Gsaxner, Zhao and
  Egger}]{radl2022avt}
\bibinfo{author}{Radl, L.}, \bibinfo{author}{Jin, Y.}, \bibinfo{author}{Pepe,
  A.}, \bibinfo{author}{Li, J.}, \bibinfo{author}{Gsaxner, C.},
  \bibinfo{author}{Zhao, F.h.}, \bibinfo{author}{Egger, J.},
  \bibinfo{year}{2022}.
\newblock \bibinfo{title}{Avt: Multicenter aortic vessel tree cta dataset
  collection with ground truth segmentation masks}.
\newblock \bibinfo{journal}{Data in brief} \bibinfo{volume}{40},
  \bibinfo{pages}{107801}.
%Type = Article
\bibitem[{Ren et~al.(2023)Ren, Wang and Zhang}]{ren2023weakly}
\bibinfo{author}{Ren, Z.}, \bibinfo{author}{Wang, S.}, \bibinfo{author}{Zhang,
  Y.}, \bibinfo{year}{2023}.
\newblock \bibinfo{title}{Weakly supervised machine learning}.
\newblock \bibinfo{journal}{CAAI Transactions on Intelligence Technology}
  \bibinfo{volume}{8}, \bibinfo{pages}{549--580}.
%Type = Inproceedings
\bibitem[{Ronneberger et~al.(2015)Ronneberger, Fischer and
  Brox}]{ronneberger2015u}
\bibinfo{author}{Ronneberger, O.}, \bibinfo{author}{Fischer, P.},
  \bibinfo{author}{Brox, T.}, \bibinfo{year}{2015}.
\newblock \bibinfo{title}{U-net: Convolutional networks for biomedical image
  segmentation}, in: \bibinfo{booktitle}{Medical Image Computing and
  Computer-Assisted Intervention--MICCAI 2015: 18th International Conference,
  Munich, Germany, October 5-9, 2015, Proceedings, Part III 18},
  \bibinfo{organization}{Springer}. pp. \bibinfo{pages}{234--241}.
%Type = Article
\bibitem[{Schneider et~al.(2012)Schneider, Rasband and
  Eliceiri}]{schneider2012nih}
\bibinfo{author}{Schneider, C.A.}, \bibinfo{author}{Rasband, W.S.},
  \bibinfo{author}{Eliceiri, K.W.}, \bibinfo{year}{2012}.
\newblock \bibinfo{title}{Nih image to imagej: 25 years of image analysis}.
\newblock \bibinfo{journal}{Nature methods} \bibinfo{volume}{9},
  \bibinfo{pages}{671--675}.
%Type = Inproceedings
\bibitem[{Selvaraju et~al.(2017)Selvaraju, Cogswell, Das, Vedantam, Parikh and
  Batra}]{selvaraju2017grad}
\bibinfo{author}{Selvaraju, R.R.}, \bibinfo{author}{Cogswell, M.},
  \bibinfo{author}{Das, A.}, \bibinfo{author}{Vedantam, R.},
  \bibinfo{author}{Parikh, D.}, \bibinfo{author}{Batra, D.},
  \bibinfo{year}{2017}.
\newblock \bibinfo{title}{Grad-cam: Visual explanations from deep networks via
  gradient-based localization}, in: \bibinfo{booktitle}{Proceedings of the IEEE
  international conference on computer vision}, pp. \bibinfo{pages}{618--626}.
%Type = Article
\bibitem[{Sun et~al.(2012)Sun, Choo and Ng}]{sun2012coronary}
\bibinfo{author}{Sun, Z.}, \bibinfo{author}{Choo, G.}, \bibinfo{author}{Ng,
  K.H.}, \bibinfo{year}{2012}.
\newblock \bibinfo{title}{Coronary ct angiography: current status and
  continuing challenges}.
\newblock \bibinfo{journal}{The British journal of radiology}
  \bibinfo{volume}{85}, \bibinfo{pages}{495--510}.
%Type = Article
\bibitem[{Tajbakhsh et~al.(2020)Tajbakhsh, Jeyaseelan, Li, Chiang, Wu and
  Ding}]{tajbakhsh2020embracing}
\bibinfo{author}{Tajbakhsh, N.}, \bibinfo{author}{Jeyaseelan, L.},
  \bibinfo{author}{Li, Q.}, \bibinfo{author}{Chiang, J.N.},
  \bibinfo{author}{Wu, Z.}, \bibinfo{author}{Ding, X.}, \bibinfo{year}{2020}.
\newblock \bibinfo{title}{Embracing imperfect datasets: A review of deep
  learning solutions for medical image segmentation}.
\newblock \bibinfo{journal}{Medical Image Analysis} \bibinfo{volume}{63},
  \bibinfo{pages}{101693}.
%Type = Article
\bibitem[{Tajbakhsh et~al.(2021)Tajbakhsh, Roth, Terzopoulos and
  Liang}]{tajbakhsh2021guest}
\bibinfo{author}{Tajbakhsh, N.}, \bibinfo{author}{Roth, H.},
  \bibinfo{author}{Terzopoulos, D.}, \bibinfo{author}{Liang, J.},
  \bibinfo{year}{2021}.
\newblock \bibinfo{title}{Guest editorial annotation-efficient deep learning:
  the holy grail of medical imaging}.
\newblock \bibinfo{journal}{IEEE transactions on medical imaging}
  \bibinfo{volume}{40}, \bibinfo{pages}{2526--2533}.
%Type = Incollection
\bibitem[{Vagenas et~al.(2023)Vagenas, Georgas and
  Matsopoulos}]{vagenas2023deep}
\bibinfo{author}{Vagenas, T.P.}, \bibinfo{author}{Georgas, K.},
  \bibinfo{author}{Matsopoulos, G.K.}, \bibinfo{year}{2023}.
\newblock \bibinfo{title}{Deep learning-based segmentation and mesh
  reconstruction of the aortic vessel tree from cta images}, in:
  \bibinfo{booktitle}{MICCAI Challenge on Segmentation of the Aorta}.
  \bibinfo{publisher}{Springer}, pp. \bibinfo{pages}{80--94}.
%Type = Article
\bibitem[{Vaswani et~al.(2017)Vaswani, Shazeer, Parmar, Uszkoreit, Jones,
  Gomez, Kaiser and Polosukhin}]{vaswani2017attention}
\bibinfo{author}{Vaswani, A.}, \bibinfo{author}{Shazeer, N.},
  \bibinfo{author}{Parmar, N.}, \bibinfo{author}{Uszkoreit, J.},
  \bibinfo{author}{Jones, L.}, \bibinfo{author}{Gomez, A.N.},
  \bibinfo{author}{Kaiser, {\L}.}, \bibinfo{author}{Polosukhin, I.},
  \bibinfo{year}{2017}.
\newblock \bibinfo{title}{Attention is all you need}.
\newblock \bibinfo{journal}{Advances in neural information processing systems}
  \bibinfo{volume}{30}.
%Type = Inproceedings
\bibitem[{Vepa et~al.(2022)Vepa, Choi, Nakhaei, Lee, Stier, Vu, Jenkins, Yang,
  Shergill, Desphy et~al.}]{vepa2022weakly}
\bibinfo{author}{Vepa, A.}, \bibinfo{author}{Choi, A.},
  \bibinfo{author}{Nakhaei, N.}, \bibinfo{author}{Lee, W.},
  \bibinfo{author}{Stier, N.}, \bibinfo{author}{Vu, A.},
  \bibinfo{author}{Jenkins, G.}, \bibinfo{author}{Yang, X.},
  \bibinfo{author}{Shergill, M.}, \bibinfo{author}{Desphy, M.}, et~al.,
  \bibinfo{year}{2022}.
\newblock \bibinfo{title}{Weakly-supervised convolutional neural networks for
  vessel segmentation in cerebral angiography}, in:
  \bibinfo{booktitle}{Proceedings of the IEEE/CVF Winter Conference on
  Applications of Computer Vision}, pp. \bibinfo{pages}{585--594}.
%Type = Inproceedings
\bibitem[{Wang et~al.(2021)Wang, Chen, Ding, Yu, Zha and Li}]{wang2021transbts}
\bibinfo{author}{Wang, W.}, \bibinfo{author}{Chen, C.}, \bibinfo{author}{Ding,
  M.}, \bibinfo{author}{Yu, H.}, \bibinfo{author}{Zha, S.},
  \bibinfo{author}{Li, J.}, \bibinfo{year}{2021}.
\newblock \bibinfo{title}{Transbts: Multimodal brain tumor segmentation using
  transformer}, in: \bibinfo{booktitle}{Medical Image Computing and Computer
  Assisted Intervention--MICCAI 2021: 24th International Conference,
  Strasbourg, France, September 27--October 1, 2021, Proceedings, Part I 24},
  \bibinfo{organization}{Springer}. pp. \bibinfo{pages}{109--119}.
%Type = Inproceedings
\bibitem[{Wang and Voiculescu(2023)}]{wang2023weakly}
\bibinfo{author}{Wang, Z.}, \bibinfo{author}{Voiculescu, I.},
  \bibinfo{year}{2023}.
\newblock \bibinfo{title}{Weakly supervised medical image segmentation through
  dense combinations of dense pseudo-labels}, in: \bibinfo{booktitle}{MICCAI
  Workshop on Data Engineering in Medical Imaging},
  \bibinfo{organization}{Springer}. pp. \bibinfo{pages}{1--10}.
%Type = Article
\bibitem[{Wasserthal et~al.(2023)Wasserthal, Breit, Meyer, Pradella, Hinck,
  Sauter, Heye, Boll, Cyriac, Yang et~al.}]{wasserthal2023totalsegmentator}
\bibinfo{author}{Wasserthal, J.}, \bibinfo{author}{Breit, H.C.},
  \bibinfo{author}{Meyer, M.T.}, \bibinfo{author}{Pradella, M.},
  \bibinfo{author}{Hinck, D.}, \bibinfo{author}{Sauter, A.W.},
  \bibinfo{author}{Heye, T.}, \bibinfo{author}{Boll, D.T.},
  \bibinfo{author}{Cyriac, J.}, \bibinfo{author}{Yang, S.}, et~al.,
  \bibinfo{year}{2023}.
\newblock \bibinfo{title}{Totalsegmentator: Robust segmentation of 104 anatomic
  structures in ct images}.
\newblock \bibinfo{journal}{Radiology: Artificial Intelligence}
  \bibinfo{volume}{5}.
%Type = Inproceedings
\bibitem[{Wei et~al.(2017)Wei, Feng, Liang, Cheng, Zhao and
  Yan}]{wei2017object}
\bibinfo{author}{Wei, Y.}, \bibinfo{author}{Feng, J.}, \bibinfo{author}{Liang,
  X.}, \bibinfo{author}{Cheng, M.M.}, \bibinfo{author}{Zhao, Y.},
  \bibinfo{author}{Yan, S.}, \bibinfo{year}{2017}.
\newblock \bibinfo{title}{Object region mining with adversarial erasing: A
  simple classification to semantic segmentation approach}, in:
  \bibinfo{booktitle}{Proceedings of the IEEE conference on computer vision and
  pattern recognition}, pp. \bibinfo{pages}{1568--1576}.
%Type = Article
\bibitem[{Weisstein(2014)}]{weisstein2014ellipse}
\bibinfo{author}{Weisstein, E.W.}, \bibinfo{year}{2014}.
\newblock \bibinfo{title}{Ellipse. from mathworld--a wolfram web resource}.
\newblock \bibinfo{journal}{From MathWorld-A Wolfram Web Resource.
  http://mathworld.wolfram.com/Ellipse.html} .
%Type = Inproceedings
\bibitem[{Wu et~al.(2022)Wu, Chen, Huang and Yue}]{wu2022weakly}
\bibinfo{author}{Wu, Q.}, \bibinfo{author}{Chen, Y.}, \bibinfo{author}{Huang,
  N.}, \bibinfo{author}{Yue, X.}, \bibinfo{year}{2022}.
\newblock \bibinfo{title}{Weakly-supervised cerebrovascular segmentation
  network with shape prior and model indicator}, in:
  \bibinfo{booktitle}{Proceedings of the 2022 International Conference on
  Multimedia Retrieval}, pp. \bibinfo{pages}{668--676}.
%Type = Inproceedings
\bibitem[{Xu et~al.(2023)Xu, Lee, Sosnovtseva, S{\o}rensen, Erleben and
  Darkner}]{xu2023extremely}
\bibinfo{author}{Xu, P.}, \bibinfo{author}{Lee, B.},
  \bibinfo{author}{Sosnovtseva, O.}, \bibinfo{author}{S{\o}rensen, C.M.},
  \bibinfo{author}{Erleben, K.}, \bibinfo{author}{Darkner, S.},
  \bibinfo{year}{2023}.
\newblock \bibinfo{title}{Extremely weakly-supervised blood vessel segmentation
  with physiologically based synthesis and domain adaptation}, in:
  \bibinfo{booktitle}{Workshop on Medical Image Learning with Limited and Noisy
  Data}, \bibinfo{organization}{Springer}. pp. \bibinfo{pages}{191--201}.
%Type = Article
\bibitem[{Zhang et~al.(2020)Zhang, Wang, Xie, Zhang, Huang, Zhang and
  Gu}]{zhang2020weakly}
\bibinfo{author}{Zhang, J.}, \bibinfo{author}{Wang, G.}, \bibinfo{author}{Xie,
  H.}, \bibinfo{author}{Zhang, S.}, \bibinfo{author}{Huang, N.},
  \bibinfo{author}{Zhang, S.}, \bibinfo{author}{Gu, L.}, \bibinfo{year}{2020}.
\newblock \bibinfo{title}{Weakly supervised vessel segmentation in x-ray
  angiograms by self-paced learning from noisy labels with suggestive
  annotation}.
\newblock \bibinfo{journal}{Neurocomputing} \bibinfo{volume}{417},
  \bibinfo{pages}{114--127}.
%Type = Article
\bibitem[{Zhang et~al.(2018)Zhang, Liu and Wang}]{zhang2018road}
\bibinfo{author}{Zhang, Z.}, \bibinfo{author}{Liu, Q.}, \bibinfo{author}{Wang,
  Y.}, \bibinfo{year}{2018}.
\newblock \bibinfo{title}{Road extraction by deep residual u-net}.
\newblock \bibinfo{journal}{IEEE Geoscience and Remote Sensing Letters}
  \bibinfo{volume}{15}, \bibinfo{pages}{749--753}.
%Type = Inproceedings
\bibitem[{Zhou et~al.(2016)Zhou, Khosla, Lapedriza, Oliva and
  Torralba}]{zhou2016learning}
\bibinfo{author}{Zhou, B.}, \bibinfo{author}{Khosla, A.},
  \bibinfo{author}{Lapedriza, A.}, \bibinfo{author}{Oliva, A.},
  \bibinfo{author}{Torralba, A.}, \bibinfo{year}{2016}.
\newblock \bibinfo{title}{Learning deep features for discriminative
  localization}, in: \bibinfo{booktitle}{Proceedings of the IEEE conference on
  computer vision and pattern recognition}, pp. \bibinfo{pages}{2921--2929}.
%Type = Inproceedings
\bibitem[{Zhou et~al.(2023)Zhou, Xu, Zhou and Tong}]{zhou2023weakly}
\bibinfo{author}{Zhou, M.}, \bibinfo{author}{Xu, Z.}, \bibinfo{author}{Zhou,
  K.}, \bibinfo{author}{Tong, R.K.y.}, \bibinfo{year}{2023}.
\newblock \bibinfo{title}{Weakly supervised medical image segmentation via
  superpixel-guided scribble walking and class-wise contrastive
  regularization}, in: \bibinfo{booktitle}{International Conference on Medical
  Image Computing and Computer-Assisted Intervention},
  \bibinfo{organization}{Springer}. pp. \bibinfo{pages}{137--147}.
%Type = Article
\bibitem[{Zhou(2018)}]{zhou2018brief}
\bibinfo{author}{Zhou, Z.H.}, \bibinfo{year}{2018}.
\newblock \bibinfo{title}{A brief introduction to weakly supervised learning}.
\newblock \bibinfo{journal}{National science review} \bibinfo{volume}{5},
  \bibinfo{pages}{44--53}.
%Type = Article
\bibitem[{Zhu et~al.(2024)Zhu, Nie and Yang}]{zhu2024tsp}
\bibinfo{author}{Zhu, Y.}, \bibinfo{author}{Nie, Z.}, \bibinfo{author}{Yang,
  X.}, \bibinfo{year}{2024}.
\newblock \bibinfo{title}{Tsp-warp-x: A novel topological shape point metric
  warping loss for fully-supervised and weakly-supervised vessel segmentation}.
\newblock \bibinfo{journal}{Authorea Preprints} .
%Type = Article
\bibitem[{Zhuang et~al.(2024)Zhuang, Chen, Yang, Kettunen and
  Wang}]{zhuang2024annotation}
\bibinfo{author}{Zhuang, M.}, \bibinfo{author}{Chen, Z.},
  \bibinfo{author}{Yang, Y.}, \bibinfo{author}{Kettunen, L.},
  \bibinfo{author}{Wang, H.}, \bibinfo{year}{2024}.
\newblock \bibinfo{title}{Annotation-efficient training of medical image
  segmentation network based on scribble guidance in difficult areas}.
\newblock \bibinfo{journal}{International Journal of Computer Assisted
  Radiology and Surgery} \bibinfo{volume}{19}, \bibinfo{pages}{87--96}.

\end{thebibliography}

% \section*{Supplementary Material}

% Supplementary material that may be helpful in the review process should
% be prepared and provided as a separate electronic file. That file can
% then be transformed into PDF format and submitted along with the
% manuscript and graphic files to the appropriate editorial office.

\end{document}